\documentclass[aps,prl,twocolumn,showpacs,10pt,preprintnumbers,nofootinbib]{revtex4-2}

\usepackage[margin=7pt,justification=centerlast]{caption}

\usepackage{amsmath}
\usepackage{amsfonts}
\usepackage{amssymb}
\usepackage{graphicx, rotating}
\usepackage{caption}
\usepackage{subcaption}
\usepackage{epstopdf}
\usepackage{epsfig}
\usepackage{latexsym}
\usepackage{mathtools}
\usepackage{MnSymbol}
\usepackage{graphicx}
\usepackage{color}
\usepackage[dvipsnames]{xcolor}
\usepackage{amsmath,amssymb}
\usepackage{hhline}
\usepackage{array,makecell}
\usepackage[utf8]{inputenc}
\usepackage{arydshln}
\usepackage{ mathrsfs }
\usepackage{orcidlink}
\usepackage{braket}
\usepackage{contour}
\usepackage{booktabs}
\usepackage{comment}

\usepackage{tikz-feynman}
\usetikzlibrary{decorations.text}
\tikzfeynmanset{warn luatex=false}

\usepackage{slashed}
\usepackage{hyperref}

\hypersetup{
    colorlinks=true,
    linkcolor=black,
    citecolor=black,
    filecolor=black,
    urlcolor=black,
    pdfborder={0 0 0}
}

\usepackage{eso-pic}
\newcommand{\reportnum}[2]{
  \AddToShipoutPictureBG*{%
    \AtPageUpperLeft{%
      \hspace{0.77\paperwidth}%
      \raisebox{#1\baselineskip}{%
        \makebox[0pt][l]{\textnormal{#2}}
  }}}%
}

\newcommand{\be}{\begin{equation}}
\newcommand{\ee}{\end{equation}}       
\newcommand{\TeV}{\,\mathrm{TeV}}

\newcommand{\gst}{g_*}
\newcommand{\mst}{m_*}
\newcommand{\eu}{\varepsilon_{u}}
\newcommand{\et}{\varepsilon_{t}}
\newcommand{\ed}{\varepsilon_{d}}
\renewcommand{\ee}{\varepsilon_{e}}

\newcommand{\Lfl}{\Lambda_{\text{fl}}}
\newcommand{\LUV}{\Lambda_{\text{UV}}}
\newcommand{\LL}{\tilde{\lambda}}
\newcommand{\tg}{\tilde{\gamma}}
\newcommand{\fp}[1]{{\left.#1\right|_{\text{fp}}}}

\newcommand{\op}{\mathcal{O}}

\reportnum{-3}{UMD-PP-025-04}

\begin{document}

\title{Partial Compositeness: from Anarchy to Symmetry}

\author{Kaustubh Agashe~\orcidlink{0009-0001-1939-3807}}
\affiliation{Maryland Center for Fundamental Physics, Department of Physics,
	University of Maryland, College Park, MD 20742, USA}
\author{Lorenzo Ricci~\orcidlink{0000-0001-8704-3545}}
\affiliation{Maryland Center for Fundamental Physics, Department of Physics,
	University of Maryland, College Park, MD 20742, USA}
\author{Raman Sundrum~\orcidlink{0009-0004-7537-5357}}
\affiliation{Maryland Center for Fundamental Physics, Department of Physics,
	University of Maryland, College Park, MD 20742, USA}

\begin{abstract}
\noindent 
Within the Composite Higgs paradigm, Partial Compositeness has emerged as an elegant mechanism for generating large flavor hierarchies such as are observed in the quark and lepton masses and mixings. This mechanism exploits the strong renormalization group effects of  the compositeness dynamics when these are {\it not} flavor-symmetric. Despite its remarkable properties, at this point it is stringently constrained  by the body of  flavor- and CP-violation tests, so that the compositeness scale must be at least $O(100)$ TeV, beyond the direct reach of proposed colliders. On the other hand, Composite Higgs theories with flavor-symmetric strong dynamics, but with realistic flavor-violating hierarchies introduced in an ad hoc manner, can extend the GIM mechanism of the standard model and thereby be far less constrained, at the edge of LHC reach and well within reach of future colliders. 
We show how the best features of both these types of dynamics can be combined if flavor-symmetries of the strong composite dynamics are emergent in the IR near the compositeness scale but absent in the far UV. In this case, flavor hierarchies can be generated by the renormalization group flow in the UV, 
 followed by an IR stage in which the dynamics flows towards accidental flavor and CP symmetries. 
  We point out how the  collider and low-energy phenomenology is significantly impacted by the IR stage. 
 Our analysis includes a discussion of the distinctive features of the small neutrino masses and their large mixings. 
\end{abstract}

\maketitle

\medskip

\section{Introduction}

The Standard Model's (SM) Higgs mechanism and Higgs boson remain among its most enigmatic features and are the focus of intense experimental efforts to probe. Its couplings are central to the hierarchical structure of particle physics that we see, both in the sense of the Hierarchy Problem related to the scale of EW symmetry breaking, $v \ll m_{\text{Planck}}$, as well as the pattern of hierarchical quark and lepton masses and mixings. The Composite Higgs beyond-SM (BSM) paradigm offers one of the boldest proposals for addressing these mysteries (see, e.g., \cite{Panico:2015jxa} for a review). The defining hypothesis is that the Higgs doublet is a composite of some as-yet unknown strongly-coupled constituents assembled at a scale $\sim m_*$. The Higgs is significantly lighter than $m_*$ because it is a pseudo-Nambu-Goldstone Boson (pNGB) of the strong dynamics.
If $m_*$ is not too far above $v$, then this compositeness solves the bulk of the Hierarchy Problem. The new strong dynamics creating the Higgs then also offers the possibility of understanding how its hierarchical Yukawa couplings emerge. Experimentally, the generic manifestation of the new physics would be the appearance of other composite excitations, also carrying SM gauge charges, with masses $\sim O(m_*)$, which may well be within collider reach. Even for collider energies $E < m_*$, off-shell composites will mediate new non-renormalizably weak interactions which can be probed at colliders or, if combined with distinctive breaking of approximate SM symmetries, within low-energy precision tests. 

{\it Partial Compositeness} \cite{Kaplan:1991dc} is a generic and important possibility: the observed particle mass-eigenstates can be superpositions of elementary particles and composite particles, where the  superposition is quantitatively determined by the strong dynamics renormalization group (RG) flow from the far UV. Phenomenologically, these superpositions determine the degree
to which different mass eigenstates participate in the new interactions. 

Beyond these general considerations, such strongly-coupled BSM theories are hard to study or even formulate microscopically. However, we can take advantage of two powerful expansion parameters to develop provisional theories.  One is the $1/N \ll 1$ expansion familiar from QCD \cite{Coleman_1985}, where $N$ is the number of colors, which can readily generalize to, and simplify, BSM strong dynamics by making the composites couple with typical strength $g_* \sim 4 \pi/\sqrt{N}$, intermediate between SM couplings and maximal strong coupling $4 \pi$. Furthermore, the simplest robust form of strong dynamics is one which is approximately conformally (scale) invariant above $m_*$, governed in large part by the non-trivial scaling dimensions $\Delta_{\op}$ of its primary composite operators $\op$. 

If only a finite subset of operators $\op$ have $\Delta_{\op} \leq \Delta_{gap}$ for some $\Delta_{gap} \gg 1$, then we can also expand in $1/\Delta_{gap} \ll 1$. Remarkably, via AdS/CFT Duality, this expansion is given precisely by AdS$_5$ effective field theory (EFT) keeping only the finite set of 5D fields dual to this subset of operators (see, e.g., the discussion in \cite{Sundrum:2011ic}). When the strong dynamics is coupled to external elementary particles/fields, the dual description is given by warped 5D EFTs of Randall-Sundrum I (RS1) type \cite{Randall:1999ee} (see, e.g., \cite{Sundrum:2005jf,Gherghetta:2010cj} for reviews). The fifth dimension is now an interval which roughly geometrizes and reflects the physics of the strong RG flow between a far UV scale $\Lambda_{\text{UV}}$ and $m_*$ \cite{Arkani-Hamed:2000ijo,Rattazzi:2000hs}. Partial compositeness is reflected in the extra-dimensional profiles of the lightest 4D modes of the 5D fields \cite{Contino:2003ve,Contino:2004vy,Agashe:2004rs}, and new strong interactions are modulated by the extra-dimensional overlap of the participating 4D particles \cite{Grossman:1999ra,Gherghetta:2000qt}.  Currently, the most explicit realistic modeling  is formulated at the level of such non-renormalizable 5D RS1 EFTs.

{\it Anarchic Partial Compositeness} (APC) is the simplest theoretically compelling realization of this dynamics, where ``anarchy'' refers to the minimal use of (approximate) symmetries. Here, SM fermion and gauge fields, each couple to different ${\cal O}$ of the strong sector, whose scaling dimensions then determine the partial compositeness of the observed lightest states through the RG flow. For a ``generic'' spectrum of scaling dimensions the EFT $< m_*$ has hierarchical structure {\it qualitatively} similar to that of the SM EFT we observe, and quantitatively can be fit to it. 
In this framework, the lightest mass eigenstates are correlated with having the smallest superposition with strong composites, so that they have the weakest new interactions in general. This greatly reduces sensitivity to the new physics within
our most powerful experimental precision low-energy tests precisely because these apply to the lightest particles \cite{Gherghetta:2000qt,Huber:2000ie,Huber:2003tu,Agashe:2004cp}.\footnote{The standard Strong CP Problem associated with neutron EDM constraints is not mitigated in this way and  requires a separate solution, such as the QCD axion mechanism with which partial compositeness is fully compatible.} Consequently, most of the precision constraints are compatible with $m_* \sim 10 - 50$ TeV, at the boundary of plausible collider reach (see, e.g., \cite{Glioti:2024hye} for updated bounds).

But dramatic improvements of electron EDM measurements in the last decades provide one of the most powerful constraints on BSM CP-violation in general, and APC in particular, for which $m_* >$ PeV, far above any contemplated collider reach! However, just mild use of leptonic symmetries of the strong dynamics relaxes the leptonic constraints back to $m_* \sim 10 - 50$ TeV \cite{Frigerio:2018uwx}. 

\begin{figure}[t!]   
    \begin{tikzpicture}[scale=1.1]
      \foreach \x/\lbl in {0/$\LUV$, 6/$\mst$} {
         \draw[line width=0.5mm,gray] (\x,-1.8) -- (\x,2);
         \node at (\x,-2){\lbl};
      }
    \node at (1.1,0.8) {\textcolor{red}{$q_L^{1,2}$}};
    \node at (4.4,0.2) {\textcolor{teal}{$u_R^3,q_L^3$}};
    \node at (0.4,-0.14) {\textcolor{blue}{$d_R^{1,2,3}$}};
    \node at (0.4,0.5) {\textcolor{blue}{$u_R^{1,2}$}};
    \node at (5,-0.3) {\textcolor{gray}{$H$}};
      
      
    \draw[thick, blue, domain=0:6, samples=200, smooth] 
    plot (\x, {(6 - \x)*(6 - \x)/24-1.1});
    \draw[thick, red, domain=0:6, samples=200, smooth] 
    plot (\x, {(6 - \x)*(6 - \x)*(6 - \x)/70-1.4});
    \draw[thick, teal, domain=0:6, samples=200, smooth] 
    plot (\x, {(\x)*(\x)/24-1});
    \draw[thick, gray, domain=0:6, samples=200, smooth] 
    plot (\x, {(\x)*(\x)*(\x)*(\x)*(\x)/2530-1.2});
    \end{tikzpicture}

  \caption{Holographic realization of APC. Chiral-zero mode profile overlaps with the Higgs are the key determinant of the SM Yukawa couplings.}
  \label{fig:CartoonAnarchy}
\end{figure}
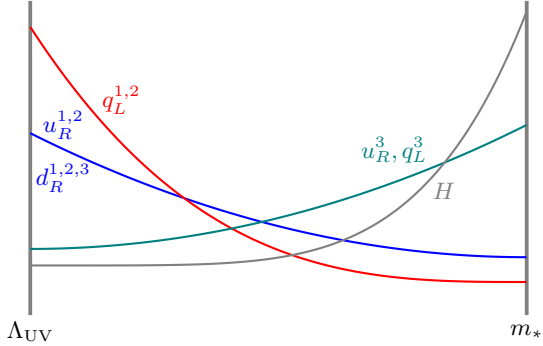

Fig.~\ref{fig:CartoonAnarchy} sketches the form of the dual 5D warped EFT structure, in particular the distinctive profiles of the different species of light 4D modes. These are determined by the 5D masses, related to the $\Delta_{\op}$ via the standard AdS/CFT dictionary. 

Given the plausibility of accidental/emergent symmetries of strongly coupled QFTs (e.g.~isospin or baryon-number within QCD, or even CP-symmetry within massless QCD) and their power to relax flavor/CP/EW precision constraints, it is important to assess how low $m_*$ can be if the strong dynamics at this scale is governed by more symmetry. 
Remarkably, maximal flavor symmetry of the strong dynamics, with explicit breaking by spurions proportional to SM Yukawa matrices \cite{Chivukula:1987py,Redi:2011zi, Barbieri:2012tu, Redi:2013pga} (Minimal Flavor Violation (MFV) \cite{DAmbrosio:2002vsn}), can relax precision constraints to $m_* \gtrsim 10$ TeV \cite{Glioti:2024hye}, now well within proposed future collider reach \cite{Golling:2016gvc,Chen:2022msz,Accettura:2023ked,Stefanek:2024kds}. A more nuanced symmetry implementation \cite{Redi:2012uj, Barbieri:2012uh,Glioti:2024hye} can even remove precision low-energy constraints as the dominant bound on $m_*$, leaving $m_* \gtrsim 5$ TeV bounded only by direct resonance searches at the LHC.

Such highly-symmetric Composite Higgs frameworks with the lowest allowed $m_*$ considerably mitigate residual EW fine-tuning and strongly motivates ongoing resonance searches at the LHC and future colliders. But Refs.~\cite{Redi:2011zi, Barbieri:2012tu, Redi:2013pga,Redi:2012uj,Glioti:2024hye,Barbieri:2012uh} leave important questions unanswered: the emergence of the requisite flavor symmetries and the 
hierarchical Yukawa matrices that explicitly break them.\footnote{See also \cite{Fitzpatrick:2007sa,Panico:2016ull,Fuentes-Martin:2022xnb,Covone:2024elw} for an incomplete list of UV realizations of Composite Higgs models that also address an explanation of flavor hierarchies or suppression of flavor violation \cite{Cheung:2007bu,Cacciapaglia:2007fw,Csaki:2008eh,Csaki:2009bb}.}  They also do not address the consistency of hierarchical charged lepton masses with the small but non-hierarchical structure of neutrino masses and mixings. In this paper, we will address how all these features can emerge from a strong RG flow from the UV in the context of partial compositeness. 

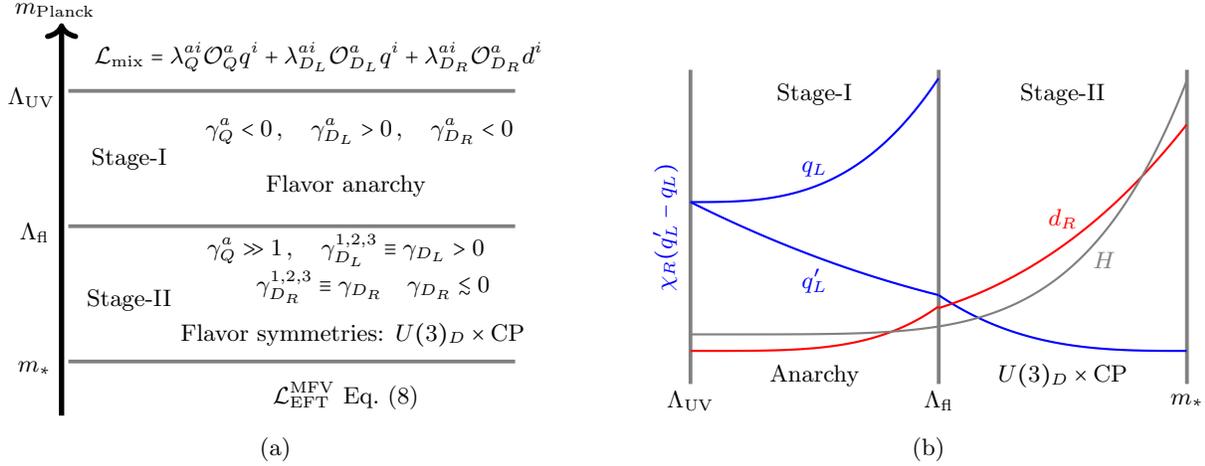
\begin{figure*}[t!]   
  \centering
  \begin{subfigure}[t]{0.48\textwidth}
    \centering
      \begin{tikzpicture}[scale=0.9]
        \draw[line width=0.5mm,gray] (-3.3,2) -- (3.3,2);
        \node at (-3.85,1.9){$\LUV$};
        \node at (-3.5,3.2){$m_{\text{Planck}}$};
        \draw[line width=0.5mm,gray] (-3.3,0) -- (3.3,0);
        \draw[line width=0.5mm,gray] (-3.3,-2) -- (3.3,-2);
        \node at (-3.8,-0.1){$\Lfl$};
        \node at (-3.8,-2.1){$\mst$};
        \node at (1,1.4){$\gamma_Q^a<0\,,\quad \gamma_{D_L}^a>0\,,\quad \gamma_{D_R}^a<0$};
        \node at (0.8,-0.35){$\gamma_Q^a\gg1\,,\quad \gamma_{D_L}^{1,2,3}\equiv\gamma_{D_L}>0$};
        \node at (1.2,-0.9){$\gamma_{D_R}^{1,2,3}\equiv\gamma_{D_R}\quad \gamma_{D_R}\lesssim0$};
        \node at (0.8,0.6){Flavor anarchy};
        \node at (0.9,-1.6){Flavor symmetries: $U(3)_D\times \mathrm{CP}$};
        \draw[->,line width=0.7mm] (-3.4,-2.8) -- (-3.4,3);
        \node at (0.8,-2.5){$\mathcal{L}_{\text{EFT}}^{\text{MFV}} $ Eq.~\eqref{Eq:MFVAns}};
        \node at (0.4,2.5){%
          $\mathcal{L}_{\text{mix}}=\lambda_Q^{ai}\mathcal{O}_Q^a q^i
          +\lambda_{D_L}^{ai}\mathcal{O}_{D_L}^a q^i
          +\lambda_{D_R}^{ai}\mathcal{O}_{D_R}^a d^i$};
        \node at (-2.4,1){Stage‑I};
        \node at (-2.4,-1.1){Stage‑II};
      \end{tikzpicture}
    \caption[]{}
    \label{fig:Cartoon:a}
  \end{subfigure}%
  \hfill
  \begin{subfigure}[t]{0.48\textwidth}
    \centering
      \begin{tikzpicture}[scale=1.1]
        \foreach \x/\lbl in {0/$\LUV$, 3/$\Lfl$, 6/$\mst$} {
          \draw[line width=0.5mm,gray] (\x,-1.8) -- (\x,2);
          \node at (\x,-2){\lbl};
        }
        \node at (1.5,-1.7){Anarchy};
        \node at (1.5,1.7){Stage‑I};
        \node at (4.5,-1.7){$U(3)_D\times \mathrm{CP}$};
        \node at (4.5,1.7){Stage‑II};
        \node at (1.5,0.8){\textcolor{blue}{$q_L$}};
        \node at (1.5,-0.55){\textcolor{blue}{$q_L'$}};
        \node[rotate=90] at (-0.3,0.1){\textcolor{blue}{$\chi_R(q_L'-q_L)$}};
        \node at (4.5,0.2){\textcolor{red}{$d_R$}};
        \node at (5,-0.3){\textcolor{gray}{$H$}};
        \draw[thick, blue, domain=0:3, samples=200, smooth]
          plot (\x,{(6-\x)*(6-\x)/24-1.1});
        \draw[thick, blue, domain=3:6, samples=200, smooth]
          plot (\x,{(6-\x)*(6-\x)*(6-\x)/40-1.4});
        \draw[thick, red, domain=0:3, samples=200, smooth]
          plot (\x,{(\x)*(\x)*(\x)*(\x)/150-1.4});
        \draw[thick, red, domain=3:6, samples=200, smooth]
          plot (\x,{(\x)*(\x)*(\x)/85-1.2});
        \draw[thick, gray, domain=0:6, samples=200, smooth]
          plot (\x,{(\x)*(\x)*(\x)*(\x)*(\x)/2530-1.2});
        \draw[thick, blue, domain=0:3, samples=200, smooth]
          plot (\x,{0.4+(\x)*(\x)*(\x)/18});
      \end{tikzpicture}
    \caption[]{}
    \label{fig:Cartoon:b}
  \end{subfigure}

 \caption{Schematic description of our symmetric model from the Ads-CFT dual perspective. Only the down‑quark sector is shown. The model can be extended to account for the up‑sector and leptons, as explained in the text. (a) RG‑flow of the symmetric model from the compositeness perspective. (b) Holographic realization of the same model within a warped 5D description.
}
\label{fig:Cartoon}
\end{figure*}

An overview of our proposal can be seen from the dual 5D perspective, sketched in Fig.~\ref{fig:Cartoon:b}.
A central feature is the presence of an intermediate  defect/``brane'' dividing the 5D ``bulk'' into two sections. These reflect two stages of RG flow, each of which is approximately conformal (close to a RG fixed point). The UV stage depicted on the left has no flavor symmetries, but the flavor-violating Yukawa hierarchies are generated here, while the IR stage depicted on the right has ``protective'' flavor symmetries (dual to 5D gauge fields) and CP symmetry. We therefore denote the juncture of these two stages as the flavor scale $\Lfl$, dual to the mid-brane,  reflecting a rapid flow or ``tumbling'' (for related idea see \cite{RABY1980373,Klebanov:2000hb}) between the two RG fixed points governing the two stages (see Fig.~\ref{fig:Cartoon:a}). From the 5D EFT viewpoint it is straightforward to introduce and stabilize the mid-brane via generalizing~\cite{Lee:2021wau,Agashe:2016rle} the Goldberger-Wise mechanism \cite{Goldberger:1999uk}. 5D masses can change significantly across this brane, and effectively this allows it to be semi-permeable. As can be seen, this primarily affects the EW doublet (``left-handed'') fermionic light modes, which have a peculiar shape. One can think of this as the composition of simpler profiles of two light modes from two 5D fields, $q_L$ and $q_L'$, the first confined to propagate in the UV section and the second propagating in the entire bulk. They are tied together by UV mass-mixing on the UV boundary to a boundary-localized field $\chi$, which below $\Lambda_{\text{UV}}$ effectively becomes a Lagrange multiplier identifying the $q_L$ and $ q_L'$ light modes.\footnote{Equivalently, we can avoid having the brane-localized fermion by choosing different boundary conditions (BC) for $q_L$ and $q_L'$. In the language of $Z_2\times Z_2'$ orbifold compactifications (see, e.g., \cite{Barbieri:2000vh}), $q$ and $q'$ have, respectively, $(++)$ and $(-+)$ BC, where 1st/2nd sign is for UV/IR boundary. Such $q$ and $q'$ can have a mass term on the UV boundary, resulting in a single chiral zero-mode housed partly in $q_L$ and partly in $q^{\prime}_L$ as depicted in Fig.~\ref{fig:Cartoon:b}.} Furthermore, notice that all the right-handed fields $u_R$ are peaked towards $\mst$ in contrast to APC (see Fig.~\ref{fig:CartoonAnarchy}).

For concreteness, in the rest of the paper we will derive the hierarchical and symmetry(-breaking) structure of our model purely from the 4D RG perspective, AdS/CFT dual to the RS 5D EFT of Fig.~\ref{fig:Cartoon}. We start in Sec.~\ref{Sec:Anarchy} where we introduce the general setup and review APC in more detail. Then, in Sec.~\ref{Sec:Symm}, we describe our symmetric model, including both quark and lepton hierarchies. We conclude in Sec.~\ref{Sec:Outlook} with comparison between the symmetry-based and APC frameworks and in particular the significant phenomenological distinctions. App.~\ref{App:ExtraDetails} provides more detail on the RG evolution of our model and in particular nuances associated to the large top Yukawa coupling. App.~\ref{App:LessSModel} expands the discussion to less-symmetric, yet not anarchic, models and App.~\ref{App:ExpCon} summarizes the current experimental constraints on the various scenarios.

\section{Anarchic Partial Compositeness}\label{Sec:Anarchy}

We start from a bottom-up Effective Lagrangian for just the SM fields,  
\begin{align}\label{Eq:GenEFT}
    {\cal L}_{\rm EFT}=\frac{m_*^4}{g_*^2}\widehat{\cal L}_{\rm EFT}\left(\frac{g_*H}{m_*},\frac{D_\mu}{m_*},\frac{\lambda^{\alpha i}_{\psi} \psi^i}{m_*^{3/2}},\ldots \right)\,,
\end{align}
where $\psi=q_L,u_R,d_R,e_R,l_L$ are the usual chiral SM multiplets, with $i=1,2,3$ denoting generation.\footnote{The ellipsis in~\eqref{Eq:GenEFT} may include multi-linear terms involving the light fields, such as bilinear interactions like $\bar{q}^i H d^j$, which we will discuss later, or four-fermion interactions. These terms are assumed to be subleading compared to those obtained directly by expanding the terms explicitly in \eqref{Eq:GenEFT}. In addition, the ellipsis encodes the formal loop dependence, i.e., terms suppressed by factors such as $g_{\text{SM}}^2/16\pi^2$, $\gst^2/16\pi^2$, or $\lambda_{\psi}^{\alpha i}[\lambda_{\psi}^{\beta i}]^*/16\pi^2$.} This is assumed to arise after integrating out composite resonances with typical masses $\sim \mst$ and interactions $\sim\gst > g_{\text{SM}}$ \cite{Giudice:2007fh}. The elementary SM fermions are assumed to linearly mix with the composite states at $\mst$ via (multiple) couplings $\lambda_{\psi}^{ai}\lesssim\gst$ $\alpha=1,2,3,\ldots$. In the cases of emergent flavor symmetries we will take these to act on the $\alpha$ index whose nature and multiplicity we will describe below. The Higgs $H$ represents the exceptionally light composite pNGB that remains in the EFT. The Higgs has an approximate shift-symmetry which is respected by $\gst$-interactions, but broken weakly by the elementary fields via gauge couplings and the $\lambda$s. The Wilson coefficients arising from expanding \eqref{Eq:GenEFT} in a Taylor series, subject to any flavor or approximate shift symmetry, are $O(1)$.\footnote{The Higgs potential requires shift-symmetry breaking but it is not directly connected to the gauge couplings and the $\lambda$s, so that it arises from these sources of shift-symmetry breaking only radiatively, naturally making the Higgs lighter than $\mst$. A further tuning is required to obtain $v< \mst/\gst$ which is the CH incarnation of the ``little hierarchy problem'', which we do not address here (see, again \cite{Panico:2015jxa}).}

Eq.~\eqref{Eq:GenEFT} at $O(\mst^0)$ reproduces the renormalizable SM and, in particular, the Yukawa couplings. For quarks,
\begin{align}\label{Eq:Yukawa}
\begin{aligned}
    Y_{u}^{ij} &\simeq \frac{(\lambda_{q}^{\alpha i})^*\, c_u^{\alpha \beta}\, \lambda_{u}^{b\beta}}{g_*}\,, &
    Y_{d}^{ij} &\simeq \frac{(\lambda_{q}^{\alpha i})^*\, c_d^{\alpha 
\beta}\, \lambda_{d}^{\beta j}}{g_*}\,,
\end{aligned}
\end{align}
where $c^{\alpha \beta}_{u/d}\sim O(1)$ but subject to any flavor symmetry constraint. Together with these symmetries the $\lambda_{\psi}^{\alpha i}$ will determine the precision phenomenological impact of the non-renormalizable terms in \eqref{Eq:GenEFT}.

The minimal and most appealing option is ``anarchy'', where there are no symmetries acting on $\alpha\equiv1,2,3$. For example the $c^{\alpha \beta}_{u,d}$ in \eqref{Eq:Yukawa} are anarchic matrices with $O(1)$ entries. By contrast the couplings to the elementary fermions at $\mst$ must be determined by RG evolution from the far UV, 
\begin{align}\label{Eq:MixingsUV}
    \mathcal{L}_{\text{mix}} = \sum_{\psi}\lambda_{\psi}^{\alpha i}\op_{\psi}^\alpha\psi^i\,.
\end{align}
We will normalize the fermionic operators $\op_{\psi}^a$ to have engineering dimension $5/2$, so that the $\lambda$s are dimensionless. The mixings are understood as functions of the RG scale, $\lambda_{\psi}^{\alpha i} \equiv \lambda_{\psi}^{\alpha i}(\mu)$. In the UV, the couplings $\lambda_{\psi}^{\alpha i}(\Lambda_{\text{UV}})$ are assumed to be $O(1)$ and anarchic, but they must be run down to the scale $\mst$, where \eqref{Eq:GenEFT} is defined. This running produces hierarchies via
\begin{align}
\frac{d \lambda^{\alpha i}_{\psi} (\mu)}{d \log \mu} = \gamma^{\alpha}_{\psi}\,\lambda^{\alpha i}_{\psi}(\mu)+\ldots \,,
\end{align}
where $\Delta_{\mathcal{O}_{\psi}^{\alpha}} \equiv \frac{5}{2} + \gamma_{\psi}^{\alpha}$, with $\Delta_{\psi}$ the scaling dimension of $\mathcal{O}_{\psi}^{\alpha}$ and the ellipsis denotes higher order terms in $\lambda$. In the simplest possibility, where all $\gamma_{\psi} >0$, this gives
\begin{align}\label{Eq:AnarcRG}
    \lambda^{\alpha i}_{\psi} (\mst)&\simeq \eta_{\psi}^{\alpha}\,\lambda_{\psi}^{\alpha i}(\LUV)   
    \,, 
    && \eta_{\psi}^{\alpha}=\left(\frac{m_*}{\LUV} \right)^{\gamma_{\psi}^{\alpha}}\,.
\end{align}
Remarkably for $\Lambda_{\text{UV}}\lesssim m_{\text{Planck}}$ and $\gamma_{\psi}^{\alpha} \sim O(0.1)$ and anarchic, \eqref{Eq:Yukawa} yields SM-like flavor hierarchies, and can readily be fit to the actual SM values. 

A subtlety arises for the top, since \eqref{Eq:AnarcRG} fails to generate a sizable top Yukawa. Instead, the heavy top must involve relevant deformations. In this case we have to be careful about higher order terms. For instance, just focusing on $\gamma_{u}^3<0$,
\begin{align}\label{Eq:RGTopAnarchy}
\frac{d \lambda^{33}_u }{d \log \mu} = \lambda_u^{33} \gamma^3_u + c\frac{(\lambda_u^{33} )^3}{\gst^2} \left(1+O\left(\frac{\lambda^2,\gst^2}{16 \pi^2} \right) \right) \,,
\end{align}
where we neglected subleading contributions. Notice that the $c\sim O(1)$ is necessarily positive \cite{Vecchi:2015fma}, thus
$\lambda_{u}^{33}$ approaches the IR fixed point 
\begin{align}\label{Eq:AnarcRGtop}
   \fp{ \lambda^{33}_{u}} = \gst \sqrt{- \gamma_u^3}\,,
\end{align}
setting $c=1$.
A parallel statement applies to $\lambda_{q}^{33}$ with $\gamma_q^3<0$. With these relevant couplings we can achieve fully realistic SM Yukawa couplings.\footnote{Given sizable $\lambda^{33}_{q,u}$ (i.e., coupling of elementary top/LH bottom quark to strong sector), it is clear that the SM top (and LH bottom) quark has a significant admixture of composite fermion interpolated by ${\cal O}…$.
Alternatively, we could assume that that the top/LH bottom quark is (purely) a massless composite of the strong sector, i.e., there is no elementary fermion at all. The two alternatives are not distinct, but are rather ``dual'' in sense of \cite{Contino:2004vy}. For simplicity/uniformity with treatment of other SM fermions (which are predominantly elementary), we will continue to use the original viewpoint with elementary top/LH bottom quark.} At the other extreme, small Dirac neutrino masses with anarchic mixings can be consistently included by considering irrelevant couplings to the strong dynamics which are bilinear in $\psi$ \cite{Agashe:2008fe}.

However, the $\mst$-suppressed effects in \eqref{Eq:GenEFT} generate a peculiar pattern of signatures, including sizable flavor and CP-violating effects, which are strongly constrained (see, e.g., \cite{Glioti:2024hye} for a recent analysis). In particular, CP violation in the quark sector requires $\mst\gtrsim O(10 \TeV)$ due to constraints from the neutron EDM, CP-violation in $\text{K}\text{-}\bar{\text{K}}$ transitions and D-meson decays. The situation is even more severe in the lepton sector, where bounds from $\mu \to e \gamma$ and the electron EDM imply $\mst\gtrsim O(10^2\text{-}10^3 \TeV)$.

\section{Symmetric models}\label{Sec:Symm}
 An economical option to relax the experimental constraints and lower $\mst$ is to construct flavor-symmetric models. The simplest is the        ``Minimal Flavor Violation'' (MFV),  defined from the bottom-up as in \eqref{Eq:GenEFT} by \cite{Redi:2011zi, Barbieri:2012tu}~\footnote{There are other realization of MFV in composite Higgs models (see, e.g., \cite{Glioti:2024hye}). For concreteness we only consider \eqref{Eq:MFVAns} but it is straightforward to generalized the main construction of the paper to other realizations.} 
\begin{align}\label{Eq:MFVAns}
    {\cal L}^{\text{MFV}}_{\rm EFT}=\frac{m_*^4}{g_*^2}\widehat{\cal L}_{\rm EFT}\left(\frac{\lambda^{ai}_{U_L} q^i}{m_*^{3/2}},\frac{\lambda^{ai}_{D_L} q^i}{m_*^{3/2}},\frac{\rho_{U_R} u^a}{m_*^{3/2}},\frac{\rho_{D_R} d^a}{m_*^{3/2}},\ldots \right)\,,
\end{align}
where $a=1,2,3$. We arrive at this structure from \eqref{Eq:GenEFT} starting with $\alpha=1,\ldots,6$, where the $\boldsymbol{\lambda}_{q,u,d}$, $6\times3$ matrices (boldface in matrix notation), are given by
\begin{align}\label{Eq:MFVAnsStruct}
\boldsymbol{\lambda}_{q} = \begin{bmatrix}
    \boldsymbol\lambda_{U_L}\\
    \boldsymbol\lambda_{D_L}
\end{bmatrix} ,\,\, \boldsymbol\lambda_{u} = \rho_{U_R}\begin{bmatrix}
     \mathbf{I}_3\\
    \mathbf{0}_3
\end{bmatrix},\,\, \boldsymbol\lambda_{d} = \rho_{D_R}\begin{bmatrix}
    \mathbf{0}_3 \\ \mathbf{I}_3
\end{bmatrix},
\end{align}
where $\boldsymbol{\lambda}_{U_L,D_L}$ are $3 \times 3$ matrices with elements $\lambda_{U_L}^{ai},\, \lambda_{D_L}^{ai}$, $\mathbf{I}_3$ and $\mathbf{0}_3$ are the identity and zero $3 \times 3$ matrices, and $\rho_{U_R,D_R}$ are complex numbers. This notation helps clarify a $U(3)_U\times U(3)_D$  broken global symmetry with spurions $\boldsymbol{\lambda}_{U_L,D_L}$ under which
\begin{align}\label{Eq:MFVGroup}
\lambda_{U_L}^{ai}q^i\in \bar{\mathbf{3}}_{U(3)_U},&& u^a\in \mathbf{3}_{U(3)_U}\,,
\end{align}
and similarly for the down sector.\footnote{With some abuse of notation, the representations $\mathbf{3}_{U(3)}$ and $\bar{\mathbf{3}}_{U(3)}$ denote fields in the fundamental and anti-fundamental of $SU(3)$, with arbitrary but opposite charges under the abelian $U(1)$ factor of $U(3)$.} Respecting the symmetry being broken by just these spurions in \eqref{Eq:Yukawa} yields 
\begin{align}\label{Eq:MFVAns2}
    Y_u^{ia}= c_u  \frac{ (\lambda_{U_L}^{ai})^* \rho_{U_R}}{\gst} \,,&& Y_d^{ia}= c_d  \frac{ (\lambda_{D_L}^{ai})^* \rho_{D_R}}{\gst}\,,
\end{align}
where $c_{u,d}$ are two $O(1)$ numbers. Charged leptons can be treated in complete analogy to down-type quarks.   

Note that since $\boldsymbol{\lambda}_{U_L,D_L}$ are proportional to the Yukawa matrices, then those are the only flavor spurion appearing in \eqref{Eq:MFVAns} (via \eqref{Eq:MFVAns2}), thereby generalizing the SM GIM mechanism. This results in a significant relaxation of experimental constraints to $\mst \gtrsim 10\TeV$, well within reach of future colliders \cite{Redi:2011zi,Barbieri:2012tu, Glioti:2024hye}. This is reviewed in App.~\ref{App:ExpCon} (see in particular Fig.~\ref{Fig:SummBounds}) for completeness.

The main goal of this work is to explain the peculiar ansatz in (\ref{Eq:MFVAns},~\ref{Eq:MFVAnsStruct}) within the top-down framework, depicted in Fig.~\ref{fig:Cartoon}. Specifically we now show that this ansatz can be achieved in analogy to the partial compositeness construction of \eqref{Eq:MixingsUV}, but with a more nuanced RG flow which is AdS/CFT-dual to the non-minimal features of Fig.~\ref{fig:Cartoon:b}. 

The two extra-dimensional segments are dual to two stages of RG flow (see Fig.~\ref{fig:Cartoon:a}):
\begin{itemize}
    \item \textbf{Stage-I} $\LUV > \mu > \Lambda_{\text{fl}}$: \emph{Anarchy}. Hierarchical mixings  $\boldsymbol{\lambda}_{U_L,D_L}$ between the elementary and composite \textit{left-handed} fields are generated by RG flow, leading to the observed SM Yukawa hierarchies by \eqref{Eq:MFVAns2}.
    \item \textbf{Stage-II} $\Lambda_{\text{fl}} > \mu > m_*$: \emph{Symmetries}. The strong dynamics possesses emergent CP and flavor symmetries.  The \textit{right-handed} elementary fields align, via RG evolution, with their composite partners so as to produce the $U(3)_U\times U(3)_D$ invariant structure $\boldsymbol{\lambda}_{u,d}$ in \eqref{Eq:MFVAnsStruct}.
\end{itemize}

For simplicity, we focus for now only on the quark sector and neglect subtleties associated to the large top Yukawa. These are discussed in App.~\ref{App:ExtraDetails}. Later, we discuss the generalization to leptons and particularly how to include neutrino masses.

In the far UV (Stage-I) the composite sector mixes to the elementary sector as 
\begin{align}\label{Eq:MixPhaseI}
\mathcal{L}_{\text{mix}} = \sum_{i,a=1,2,3} &\Big( \lambda_Q^{ai} \op^a_Q q^i +\lambda_{U_L}^{ai} \op_{U_L}^a q^i +\\ &\lambda_{U_R}^{ai} \op^{a}_{U_R} u^i+\lambda_{D_L}^{ai} \op^{a}_{D_L} q^i+\lambda_{D_R}^{ai} \op^{a}_{D_R} d^i\Big)\,.\notag
\end{align}
Note that the SM doublets $q$ mix with $\op_Q^a,\, \op_{U_L}^a,\,\op_{D_L}^a$ in this stage but $\boldsymbol{\lambda}_{Q}$ does not appear in the $\mst$-scale EFT in \eqref{Eq:MFVAns}. Yet, we will see later on that $\boldsymbol{\lambda}_{Q}$ participates in the RG for $\boldsymbol{\lambda}_{U_L,D_L}$.  At $\Lambda_{\text{UV}}$ no flavor and CP symmetries are present ($\LUV\lesssim m_{\text{Planck}}$) and the mixings $\lambda(\LUV)$ are complex and anarchic $O(1)$. Furthermore the Stage-I strong dynamics has no flavor symmetries. 

In the following we will just focus on the operators relevant for down-type Yukawa couplings. The scaling dimensions of the composite operators, $\Delta_{\op_{\psi}^a} \equiv \frac{5}{2}+\gamma_{\psi}^a$, are taken in the ranges
\begin{align}
    \text{Stage-I:} && \begin{cases}
        -1 < \gamma_Q^1<\gamma_Q^2<\gamma_Q^3 < 0  \\ 
        \gamma_{D_L}^1>\gamma_{D_L}^2> \gamma_{D_L}^3 > 0 \\
        -1 < \gamma_{D_R}^1<\gamma_{D_R}^2<\gamma_{D_R}^3 < 0 
    \end{cases}
\end{align}

To describe the RG, we now consider, for simplicity, a single family, say $a = i = 1$. We report in App.~\ref{App:ExtraDetails} the straightforward generalization. The RG equations read
\begin{align}\label{Eq:RGvolutionPhaseI}
\begin{cases}
    \dfrac{d \lambda_{D_L}^{11}}{d \log \mu} = \gamma_{D_L}^{1} \lambda_{D_L}^{11} + \dfrac{\lambda_{D_L}^{11} |\lambda_Q^{11}|^2}{g_*^2} \\[10pt]
    \dfrac{d \lambda_{Q}^{11}}{d \log \mu} = \gamma_Q^1 \lambda_{Q}^{11} + \dfrac{\lambda_{Q}^{11} | \lambda_Q^{11}|^2}{g_*^2}
\end{cases}
\end{align}
where we neglected higher-order contributions in $(\lambda_{D_L}^{11})^3$ since these are suppressed in the IR. Indeed the coupling $\lambda_q^{11}$ in the second equation of \eqref{Eq:RGvolutionPhaseI} approaches the fixed point in the IR
\begin{align}\label{Eq:FixPointq}
|\fp{\lambda_Q^{11}}|^2= -\gst^2 \gamma_{Q}^1\,.
\end{align}
Inserting this in
\eqref{Eq:RGvolutionPhaseI}, gives
\begin{align}
\begin{aligned}
    &\frac{d \lambda_{D_L}^{11}}{d \log \mu} = \gamma_{D_L}^1 \lambda_{D_L}^{11} - \lambda_{D_L}^{11} \gamma_Q^{1}\\&\Rightarrow \lambda_{D_L}^{11}(\mu)=\lambda_{D_L}^{11}(\LUV)\left(\frac{\mu}{\LUV} \right)^{\gamma_{D_L}^1 - \gamma_Q^1}\,
\end{aligned}
\end{align}
where, for simplicity and robustly, we assumed $\lambda_{Q}^{11}$ to have reached the fixed point very quickly.\footnote{Indeed, as we explain in App.~\ref{App:ExtraDetails}, we can estimate $\lambda^{11}_{q}(\mu)\sim \fp{\lambda^{11}_{q}}\left(1+O\left(\delta_{\mu}^{-2\gamma_q^1}\right)\right)$ with $\delta_{\mu} =  \mu/\Lambda_{\text{UV}}$.}
Generalizing to all three generations we find
\begin{align}
    \lambda^{ai}_{D_L}(\Lfl\lesssim\mu \ll\LUV) = \eta_Q^a(\mu)\lambda^{ai}_{D_L}(\LUV) \eta_{D_L}^i(\mu)\,,
\end{align}
with
\begin{align}
    \eta_Q^a(\mu) = \left(\frac{\mu}{\LUV} \right)^{-\gamma_Q^a} \,, \,\,\,\,\, 
    \eta_{D_L}^i(\mu) = \left(\frac{\mu}{\LUV} \right)^{\gamma_{D_L}^i} \,.
\end{align}

The RH mixings are driven by a relevant deformation and they flow toward to a fixed point, similarly to~\eqref{Eq:AnarcRGtop}. In particular, at this stage $\lambda_{d}^{ai}(\mu) \lesssim \gst$. 

At the ``flavor scale'' $\Lfl$ ($\LUV \gg \Lfl \gg \mst$), the dynamics is assumed to undergo a sudden and significant transition from flavor anarchy of the strong dynamics toward emergent flavor symmetries. Above and below this transition the dynamics is taken to be governed by distinct RG fixed points, close to distinct CFTs. In particular the spectrum of scaling dimensions of low-lying operators is different. We will see that precision flavor tests constrain {$\Lfl\gtrsim (10^4-10^5)\gst\TeV $. In 5D this threshold can be robustly modeled as an intermediate brane, separating the two ``slices'' of approximately AdS$_5$ geometry. The matching coefficients of the RG at this threshold are represented by brane-localized couplings. 

Below the transition the dynamics are assumed to be near a new IR CFT. This is assumed to possess an accidental $U(3)_U\times U(3)_D$ global symmetry, with nearly marginal operators $\op_{U_{L,R},D_{L,R}}$, but no longer including $\op_Q$:
\begin{align}
    \text{Stage-II}:&& \begin{aligned} &\op^a_{U_{L,R}} \in \bar{\mathbf{3}}_{U(3)_U}\,, \quad \op^a_{D_{L,R}} \in \bar{\mathbf{3}}_{U(3)_D}\,, \\
     &\gamma_{D_L}^{1,2,3} \equiv \gamma_{D_L}>0\,, \quad
        -1 < \gamma_{D_R}^{1,2,3} \equiv \gamma_{D_R} < 0\,.
        \end{aligned}
\end{align}
Additionally the strong dynamics below $\Lfl$ is assumed to be accidentally $\text{CP}$-invariant to protect from large effects in nEDM \cite{Glioti:2024hye}. The strong dynamics at or above $\Lfl$ might also be CP-invariant or not, see \ref{Sec:Outlook} for implications.

The RG flow in Stage-II is 
\begin{align}\label{Eq:RGEvolutionPhaseII}
\begin{cases}
    \dfrac{d \lambda_{D_L}^{ai}}{d \log \mu} = \gamma_{D_L} \lambda_{D_L}^{ai}  \\[10pt]
    \dfrac{d \lambda_{D_R}^{ai}}{d \log \mu} = \gamma_{D_R} \lambda_{D_R}^{ai} + \dfrac{\lambda_{D_R}^{aj} [ \lambda_{D_R}^{\dagger}\lambda_{D_R}]^{ji}}{g_*^2}
\end{cases}
\end{align}
The RH mixings $\lambda_{D_R}^{ai}$, similarly to \eqref{Eq:RGTopAnarchy}, approach the diagonal and universal fixed point
\begin{align}\label{Eq:DiagonalRH}
    \boldsymbol{\lambda}_{D_R}(\mst) \simeq \fp{\boldsymbol{\lambda}_{D_R}}=\gst \sqrt{-\gamma_{D_R}}\,\mathbf{I}_3\,.
\end{align} 
The previous equation matches with \eqref{Eq:MFVAnsStruct}: we can identify $\rho_{D_R} = \gst \sqrt{-\gamma_D}$.

The linear RG for $\lambda_{D_L}$ in \eqref{Eq:RGEvolutionPhaseII} results in a universal rescaling, namely
\begin{align}\label{Eq:LHMixmst}
    \lambda^{ai}_{D_L}(\mst) = \left(\frac{\mst}{\Lfl} \right)^{\gamma_{D_L}} c^{ab}_{D_L} \, \eta^b_{Q}(\Lfl) \, \lambda^{bi}_{D_L}(\LUV) \, \eta^i_{D_L} (\Lfl)\,.
\end{align}
Here, $c^{ab}_{D_L}$ is an anarchic $O(1)$ matrix arising from flavor non-symmetric matching of the couplings between the I and II Stage at $\Lfl$ (again, dual to couplings localized on the intermediate brane). 

The previous expression can be written as
\begin{align}
    \lambda_{D_L}^{ai}= \left(\frac{\mst}{\Lfl} \right)^{\gamma_{D_L}}  \,U_{D}^{ab}\, \eta^b_{Q} (\Lfl) \lambda'^{bi}_{D_L}(\LUV) \, \eta^i_{D_L}(\Lfl) \,,
\end{align}
where $U_{D}^{ab}$ is unitary and we have reabsorbed $c_{D_L}^{ab}$ in the anarchic matrix $\boldsymbol{\lambda}'_{D_L}$.\footnote{In particular via rotation $U_D$ we can put $c^{ab}_{D_L}$ into a lower triangular form $\hat{c}_{D_L}^{ab}$. Then we use $\hat{c}_{D_L}^{ab}\eta_{Q}^b(\Lfl) \lambda_{D_L}^{bi}\simeq \eta_Q^b(\Lfl)\lambda^{'bi}_{D_L}(\Lfl)$ with $\lambda^{'bi}_{D_L}(\Lfl)$ another anarchic matrix.} With the rotation $\op_{D_{L,R}}^a\rightarrow \op_{D_{L,R}}^b (U_D^{ab})^{\dagger}$ and $d^i \rightarrow U_D^{ij}d_j$ we can remove $U_{D_L}$ from the previous expression still keeping \eqref{Eq:DiagonalRH}. In this new basis, we have
\begin{align}\label{Eq:MixIRDown}
    \lambda^{ai}_{D_L}(\mst) = \left(\frac{\mst}{\Lfl} \right)^{\gamma_{D_L}}\eta^b_{Q}(\Lfl){\lambda}^{'bi}_{D_L}(\LUV)\eta^i_{D_L}(\Lfl) \,,
\end{align}
The previous equation shows that plugging $\lambda_{D_L}$ in \eqref{Eq:MFVAns2} yields the same attractive pattern of hierarchical Yukawa couplings found in APC by plugging \eqref{Eq:AnarcRG} in \eqref{Eq:Yukawa}. In the up-sector, the presence of a relevant deformation associated to the large top Yukawa, generates a different structure than APC with suppressed mixing angles. Yet, the down-sector will dominate the contributions to the physical CKM angles, as we discuss in App.~\ref{App:ExtraDetails}

Notice that, as we pointed out before, in Stage-I, the theory is essentially anarchic, allowing for generic flavor-violating effects. Given that $\lambda_{Q, U_R, D_R}^{ai} (\Lfl)\lesssim \gst$ are not suppressed at this Stage, we expect $\Delta F = 2$ four-fermion operators suppressed by $\gst^2/\Lfl^2$. The coefficients of some of these operators, namely the one involving right-handed fields, will be further suppressed by the RG flow in Stage-II in line with (\ref{Eq:RGEvolutionPhaseII},~\ref{Eq:DiagonalRH}). Yet, the coefficients of four-fermion operators made purely out of $q^i$ have no strong RG suppression since the $\op_Q^a$ are not relevantly coupled to the strong dynamics in Stage-II. Bounds on $\Delta S = 2$ transitions \cite{Bona:2024bue} of this form then imply $\Lfl \gtrsim 10^5\text{--}10^6 \,\gst\, \TeV$.

A similar procedure can be applied to the up sector, though again a subtlety arises in connection with the top quark. In order to generate a sizable Yukawa, $\gamma_{U_L}^3$, in Stage-I and/or $\gamma_{U_L}$ in Stage-II must be sufficiently negative. This generalization of \eqref{Eq:RGvolutionPhaseI} is straightforward and we report the details in App.~\ref{App:ExtraDetails}. 

The benchmark values $\mst= 10 \TeV$ and $\gst=3 $, at the boundary of the allowed parameter space (see App.~\ref{App:ExpCon} for details),
together with $\,\,\Lambda_{\text{UV}}= m_{\text{Planck}}=10^{16}\TeV$, $\Lfl= 10^{16}\TeV$ and 
\begin{align}\label{Eq:Benchmark}
\small
&\begin{aligned}
\text{Stage-I: }
&\gamma_{U_L}^a \sim (0.2,\, 0.05,\,-0.05),\, 
-\gamma_Q^a \sim (0.4,\,0.3,\,0.05),\,\\
&\gamma_{D_L}^a \sim (0.1,\, 0.05,\, 0.05),\,
\,\,\,\,-\gamma_{U_R,D_R}^a \sim O(0.1)
\end{aligned}\notag\\
&\begin{aligned}
\text{Stage-II: }
\gamma_{U_L} \sim -0.15,\ 
\gamma_{D_L} \sim 0.05,\\
\gamma_{U_R} \sim -0.5,\ 
\gamma_{D_R} \sim -0.6
\end{aligned}
\end{align}
can readily be fit to Yukawa matrices, matching the real world quark masses and mixings for anarchic $O(1)$ $\boldsymbol{\lambda}(\Lambda_{\text{UV}})$ and $c_{U_L,D_L}^{ab}$.\footnote{We have numerically checked, by solving the RG equations at $O(\lambda^3)$, that these values generate Yukawa couplings and mixing angles in the ballpark of the observed SM values, starting from $O(1)$ mixings $\lambda(\Lambda_{\text{UV}})$ and $O(1)$ matching matrices $c^{ab}$ at $\Lfl$. These values are consistent with the analytic expressions in App.~\ref{App:ExtraDetails}.}
 As can be seen these $\gamma$ are consistent with themselves being anarchic and $\sim O(0.1)$.

\textbf{\textit{Leptons.--}} The previous construction can be generalized to include leptons. In particular, we extend \eqref{Eq:MFVAns} to include leptonic arguments
\begin{align}\label{Eq:MFVLeptEFT}
    \mathcal{L}_{\text{EFT}}^{\text{MFV}} \supset \frac{\mst^4}{\gst^2}\,\hat{\mathcal{L}}_{\text{EFT}}\left( \frac{\LL_{D_L}^{ai}l^i}{\mst^{3/2}},\frac{\tilde{\rho}_{D_R}e^a}{\mst^{3/2}},\frac{\gst^2\tilde{\mathcal{Y}}_{\nu}^{ij}\bar{l}^iH^C\nu^j}{\mst^4},\ldots\right)\,,
\end{align}
with $H^C \equiv i \sigma^2 H^*$, the usual conjugate of the $H$ doublet. Here, the first two terms describe linear mixings responsible for generating hierarchical charged lepton masses, in complete analogy with the down-quark sector. We have also included a gauge invariant term bilinear in leptons to generate a Dirac mass for neutrinos, where $\nu^i$ denote the three right-handed neutrinos. Throughout the rest of the text, we include a tilde (e.g., $\LL$) to distinguish lepton-sector quantities from their quark-sector counterparts. In \eqref{Eq:MFVLeptEFT} we could have included analogous bilinear terms unrelated to neutrinos. We will see that all such $\tilde{\mathcal{Y}}$ can flow to be very small but anarchic, so that they will only be important when other Yukawa contributions (from linear mixings) are highly suppressed, as is however the case only for neutrinos, in the spirit of \cite{Agashe:2008fe}. We thus arrive at 
\begin{align}\label{Eq:LeptYuk}
   Y_e^{ia} \simeq c_e \frac{(\LL^{ai}_{D_L})^* \tilde{\rho}_{D_R}}{\gst}\,,&& Y_{\nu}^{ij} \simeq \tilde{\mathcal{Y}}_{\nu}^{ij}\,,
\end{align}
so that the charged leptons can be hierarchical consistently with small but anarchic neutrino masses, as observed in the data.

The origin of \eqref{Eq:MFVLeptEFT} begins in the UV with
\begin{align}\label{Eq:MixPhaseILept}
\mathcal{L}_{\text{mix}}(\Lambda_{\text{UV}})&\supset \sum_{i,a=1,2,3} \Big( \LL_L^{ai} \tilde{\op}^a_L l^i +\LL_{U_L}^{ai} \tilde{\op}_{U_L}^a l^i +\\ &\LL_{U_R}^{ai}\tilde{\op}_{U_R}^a \nu^i + \LL_{D_L}^{ai} \tilde{\op}^{a}_{D_L} l^i+ \LL_{D_R}^{ai} \tilde{\op}^{a}_{D_R} e^i\Big)\,,\notag
\end{align}
with $\tilde{\op}_L^a,\,\tilde{\op}^a_{U_L,D_L},\,\tilde{\op}^a_{U_R,D_R}$ the composite operators.\footnote{Bilinear interactions of the form $\bar{l}\,\op_H^C\,\nu$, with $\op_H^C$ the conjugate of the operator interpolating the Higgs, are also generated at $\LUV$ and can be included in \eqref{Eq:MixPhaseILept}. However, these turn out to be suppressed by the RG.} All the $\tilde{\lambda}$ are anarchic and $O(1)$. In Stage-I we assume $-1<\tilde{\gamma}_{L,U_R, D_R}^{1,2,3}<0$, $\tilde{\gamma}_{U_L,D_L}^{1,2,3}>0$ in analogy to the quark-sector. 

At the transition at $\Lfl$ we match onto Stage-II as we did for quarks. However, we will now include irrelevant effects bilinear in the elementary fields:
\begin{align}\label{Eq:BilinearsNeut}
    \mathcal{L}_{\text{mix}}(\Lfl) \supset \tilde{\mathcal{Y}}^{ij}_{\nu}\, \bar{l}^i\, \op_H^C\, \nu^j+\tilde{\mathcal{Y}}^{ij}_{e}\, \bar{l}^i\, \op_H\, e^j\,,
\end{align}
where $\op_H^{(C)}$ denotes the (conjugate) operator interpolating the composite Higgs, normalized $\op_H$ to have engineering dimension $1$. Assuming for simplicity that $\LL_{L}^{ai}(\Lfl) \sim \gst$, we estimate $\tilde{\mathcal{Y}}^{ij}_{\nu}(\Lfl)$ as an anarchic matrix with entries of order $O(\gst)$.

In Stage-II, the theory develops emergent global $U(3)_{N}\times U(3)_E$ symmetry, so that:
\begin{equation}
\begin{aligned}
    &\tilde{\op}^a_{U_L, U_R} \in \bar{\mathbf{3}}_{U(3)_N}\,, \quad
    \tilde{\op}^a_{D_L, D_R} \in \bar{\mathbf{3}}_{U(3)_E}\,, \\
    &\tg_{U_L, D_L}^{a=1,2,3} \equiv \tg_{U_L, D_L} > 0\,, \quad
    \tg_{U_R, D_R}^{a=1,2,3} \equiv \tg_{U_R, D_R} < 0\,.
\end{aligned}
\end{equation}

Paralleling the quark sector, in Stage-II we have
\begin{align}\label{Eq:LinearTermsLeptons}
    \tilde{\lambda}_{U_L}^{ai}(\mst) &\sim \left(\frac{\mst}{\Lfl} \right)^{\tg_{U_L}} 
     \left( \frac{\Lfl}{\LUV}\right)^{\tg_{U_L}^i - \tg_L^a} 
      \,\LL^{ai}_{U_L}(\LUV)\,,\\
      \tilde{\lambda}_{D_L}^{ai}(\mst) &\sim \left(\frac{\mst}{\Lfl} \right)^{\tg_{D_L}} 
     \left( \frac{\Lfl}{\LUV}\right)^{\tg_{D_L}^i - \tg_L^a} 
      \,\LL^{ai}_{D_L}(\LUV)\,.
\end{align}

We now focus on the bilinear terms. For instance $\tilde{\mathcal{Y}}_{\nu}^{ij}$ runs as 
\begin{align}\label{Eq:RGvolutionBil}
\begin{cases}
    \dfrac{d \tilde{\mathcal{Y}}_{\nu}^{ij}}{d \log \mu} =  
    \gamma_H\, \tilde{\mathcal{Y}}_{\nu}^{ij}
    + \tilde{\mathcal{Y}}_{\nu}^{ik} \dfrac{[\LL_{U_R}^{\dagger} \LL_{U_R}]^{kj}}{\gst^2} \\[10pt]
    \dfrac{d \LL_{U_R}^{ai}}{d \log \mu} = 
    \tilde{\gamma}_{U_R}\, \LL_{U_R}^{ai}
    + \dfrac{\LL_{U_R}^{aj} \left[ \LL_{U_R}^{\dagger} \LL_{U_R} \right]^{ji}}{\gst^2} 
\end{cases}
\end{align}
where $\Delta_H= 1+\gamma_H$ is the scaling dimension of $\op_H$. We consider again the solution where the right-handed mixings in the second equation of \eqref{Eq:RGvolutionBil} reach the fixed point $\fp{\boldsymbol{\LL}_{U_R}^{ai}} \simeq \gst \sqrt{-\tilde{\gamma}_{U_R}}\,\mathbf{I}_3$. Substituting this in the first equation we get
\begin{align}\label{Eq:BilinearNeutrino}
    \tilde{\mathcal{Y}}_{\nu}^{ij} (\mst) \sim  \left( \frac{\mst}{\Lfl}\right)^{\gamma_H -\tg_{U_R}}\tilde{\mathcal{Y}}_{\nu}^{ij} (\Lfl)\,,
\end{align}
and similarly 
\begin{align}\label{Eq:Bilinearelectron}
    \tilde{\mathcal{Y}}_{e}^{ij} (\mst) \sim  \left( \frac{\mst}{\Lfl}\right)^{\gamma_H -\tg_{D_R}}\tilde{\mathcal{Y}}_{e}^{ij} (\Lfl)\,.
\end{align}

The final Yukawa couplings are the sum of the bilinear contributions (\ref{Eq:BilinearNeutrino},~\ref{Eq:Bilinearelectron}) and the one arising from linear terms $(\tilde{\lambda}_{U_R}^{ai})^* \tilde{\rho}_{U_R}/\gst$ and $(\tilde{\lambda}_{D_R}^{ai})^* \tilde{\rho}_{D_R}/\gst$.

Note that taking $\Lfl\gtrsim 10^{6}\TeV$ and $\mst\sim O(10\TeV)$, for $\gamma_H\gtrsim 1$~\footnote{This ensures that the gauge invariant scalar $\op_H^C\op_H$ is not a relevant operator (in a large-N approximation for the strong dynamics in which its dimension is approximately twice that of $\op_H$) that would reintroduce a hierarchy problem.} and $-\tilde{\gamma}_{D_L}\gtrsim O(0.1)$ the bilinear interaction $\tilde{\mathcal{Y}}_e^{ij}$ is at most comparable to the electron Yukawa. Thus bilinear terms are expected to be subdominant for charged lepton, and analogously quark, hierarchies.

Turning to neutrinos the data does not support a hierarchical structure so we need the anarchic bilinear contribution of \eqref{Eq:BilinearNeutrino} to dominate over $\tilde{\lambda}_{U_R}^{ai} \tilde{\rho}_{U_R}/\gst$. This robust possibility holds for $\tilde{\gamma}_{U_L}-\tg_{L}^a+ \tg_{U_L}^i\gtrsim\gamma_H-\tilde{\gamma}_{U_R}$ for the choice of an intermediate scale $\Lfl\sim \sqrt{\Lambda_{\text{UV}} \mst}$. Furthermore for \eqref{Eq:BilinearNeutrino} to match the data on neutrino masses we need $\gamma_H-\tilde{\gamma}_{U_R}\sim 2.5$, assuming $\mst/\Lfl\sim10^{-5}$. Consistently this requires, for instance, $\tg_{U_L},\tg_{U_L}^{a}\gtrsim 1$ to suppress the bilinear contribution.

\section{Outlook}\label{Sec:Outlook}

 We have developed a framework of Minimal Flavor Violation (MFV) for the Composite Higgs scenario which emerges in two stages of RG evolution from the far UV. In the first stage at the highest energies the dynamics is flavor-anarchic, but evolves to generate realistic flavor hierarchies for the effective SM Yukawa couplings, while in the second stage the dynamics flows towards accidental flavor (generational) $U(3)$ symmetries, broken only by the Yukawa matrices. By the Higgs compositeness scale $m_*$, this realizes the proposal of \cite{Redi:2011zi, Barbieri:2012tu,Redi:2013pga,Redi:2012uj,Glioti:2024hye} in which the SM GIM mechanism is extended to the host of compositeness effects, so that these can robustly evade low-energy precision CP-violation and flavor-violating tests for $m_* \sim 10$ TeV. 

Obviously, for phenomenological purposes it is important to ask just how low $m_*$ can be, given this type of framework. We have discussed briefly in App.~\ref{App:LessSModel} how to generalize to the less symmetric variant (for the up-type right-handed quarks) in our two-stage evolution, in which $m_* \sim 5$ TeV is possible, at the edge of LHC sensitivity. 

We also described the generation of hierarchies among the charged leptons, while obtaining (Dirac) neutrinos with much smaller but anarchic mass matrices. The neutrino mixing can be either CP-conserving or significantly CP-violating depending on whether the intermediate ``flavor'' scale separating the two stages of strong dynamics preserves CP or not, either of which is allowed. 

There is also a sharp contrast in the phenomenology of Composite Higgs between the anarchic and flavor-symmetric variants. The anarchic Composite Higgs framework possesses an imperfect realization of the GIM mechanism, and therefore is strongly constrained by low energy CP and flavor tests, generally bounding $\mst>10-50$ TeV but with the electron EDM constraint giving $m_* >$ PeV \cite{Frigerio:2018uwx}. If CP-violation is restricted to the quark sector in the UV, then the electron EDM constraint is eliminated, and plausible future  colliders could probe the composite physics. Because the heaviest SM particles in are the most (partially) composite, any composites produced at colliders decay primarily into these heaviest particles: tops, Higgs, W, Z. Because the valence quarks and leptons at any collider have negligible partial compositeness, production of composites is mostly mediated by SM gauge bosons.  On the other hand, flavor-symmetric Composite Higgs allows significantly lower $m_* \sim 5-10\TeV$, but now {\it all} generations  of (right-handed) fermions have significant partial compositeness. Therefore valence quarks and leptons at colliders can directly couple to and create the composites, and these will decay into all generations. Indeed, the tightest bounds on $m_*$ currently arise from quark compositeness bounds from the LHC. At future muon or electron colliders, the leptonic compositeness opens up powerful and exciting new channels, with significantly higher sensitivity to $\mst$ than anarchic models, which we leave for further study.  
 
\subsection*{Acknowledgments}

We would like to thank R. Rattazzi and L. Vecchi for useful discussions. This work is supported by NSF Grant No.~PHY-2210361 and by the Maryland Center for Fundamental Physics. 

\appendix
\section{From anarchy to MFV}\label{App:ExtraDetails}
In this Appendix we discuss in more detail the RG evolution that leads to the phenomenological Lagrangian in (\ref{Eq:MFVAns}, \ref{Eq:MFVAnsStruct}). 

The starting point in the UV is \eqref{Eq:MixPhaseI}, which involves the three elementary quark doublets $q^i$ coupled a total of 9 doublets operators. These operators (and their corresponding mixings) are conveniently grouped into three sets: $\mathcal{O}_Q^a$, $\mathcal{O}_{U_L}^a$, and $\mathcal{O}_{D_L}^a$, with $a = 1, \ldots, 3$. Focusing on the LH fields, the full RG at cubic order in $\lambda$ can be written as
\begin{align}\label{Eq:GenRGtop}
\begin{cases}
    \dfrac{d \lambda_{Q}^{ai}}{d \log \mu} = \gamma_{Q}^a \lambda_{Q}^{ai} + \lambda_{Q}^{aj} \dfrac{\left[\sum_{\chi=Q,U_L,D_L}\lambda_{\chi}^{\dagger}\lambda_{\chi} \right]^{ji}}{g_*^2} \\
    \dfrac{d \lambda_{U_L}^{ai}}{d \log \mu} = \gamma_{U_L}^a \lambda_{U_L}^{ai} + \lambda_{U_L}^{aj} \dfrac{\left[\sum_{\chi=Q,U_L,D_L}\lambda_{\chi}^{\dagger}\lambda_{\chi} \right]^{ji}}{g_*^2} \\
    \dfrac{d \lambda_{D_L}^{ai}}{d \log \mu} = \gamma_{D_L}^a \lambda_{D_L}^{ai} + \lambda_{D_L}^{aj} \dfrac{\left[\sum_{\chi=Q,U_L,D_L}\lambda_{\chi}^{\dagger}\lambda_{\chi} \right]^{ji}}{g_*^2}
\end{cases}
\end{align}
where we have reabsorbed $O(1)$ coefficients in the cubic terms (see \eqref{Eq:AnarcRGtop}) in the definitions of the $\lambda$s.

We assume the following scaling dimensions 
\begin{align}
\begin{aligned}
\text{Stage-I:}&&
\begin{cases}
 -1<\gamma_Q^{1}<\gamma_Q^{2}<\gamma_{U_L}^3<\gamma_Q^{3}<0\\
\gamma_{D_L}^1>\gamma_{D_L}^2>\gamma_{D_L}^3>0\\
    \gamma_{U_L}^1 > \gamma_{U_L}^2>0
    \end{cases}
\end{aligned}
\end{align}
where we have used the freedom to rearrange the indices $a$ to achieve this ordering. Note the requirement $\gamma_{U_L}^3<\gamma_Q^3$, needed to generate a sizable top Yukawa. 

To begin with, we solve analytically \eqref{Eq:GenRGtop}, making suitable approximations as follows. We first focus on the most relevant interaction, associated with $\op_{Q}^{\,1}$, namely $\lambda_Q^{1i}$, and neglect all other couplings $\lambda_{Q}^{2i},\,\lambda_{Q}^{3i},\,\lambda_{U_L,D_L}^{ai}$ for the purpose of solving the RG for $\lambda_Q^{1i}$. As we will show later, the subset of these couplings that could potentially contribute to the running of $\lambda_Q^{1i}$ is in fact suppressed by their own RG equations. Within this approximation, the RG for $\lambda_Q^{1i}$ simply reads
\begin{align}
    \frac{d \lambda_{Q}^{1i}}{d \log \mu} = \gamma^{1}_Q\lambda_Q^{1i} + \lambda_Q^{1j}(\lambda_Q^{j\,1})^* \lambda_{Q}^{1i}\,,
\end{align}
with solution 
\begin{align}\label{Eq:RG1Gamma}
    \lambda_Q^{1i}(\mu) = \frac{\lambda_Q^{1i}(\Lambda_{\text{UV}})}{ \sqrt{\delta_{\mu}^{-2\gamma_Q^1}-\frac{||\lambda_Q^1(\Lambda_{\text{UV}})||^2}{\gst^2\,\gamma_Q^1}\left(1-\delta_{\mu}^{-2\gamma_Q^1}\right)}}\,,
\end{align}
where $\delta_{\mu} \equiv \mu/\Lambda_{\text{UV}}$ and
\begin{align}
    ||\lambda_Q^{1}(\mu)|| =\sqrt{\sum_{i=1,2,3} \left( \lambda_{Q}^{1i}(\mu)\right)^2}\,.
\end{align}
In the IR, \eqref{Eq:RG1Gamma} reaches the fixed point
\begin{align}
 \fp{\lambda_Q^{1i}} = \gst \sqrt{-\gamma^1_Q} \frac{\lambda_Q^{1i} (\Lambda_{\text{UV}})}{||\lambda_Q^{1}(\Lambda_{\text{UV}})||}\,,
\end{align}
and by a rotation of the elementary quarks $q^i$ we can set 
\begin{align}\label{Eq:FP1}
   \fp{ \lambda_{Q}^{11}}=  \gst \sqrt{-\gamma_Q^1}\,,&& \fp{\lambda_{Q}^{12} }=\fp{\lambda_{Q}^{13}}= 0\,.
\end{align}
Note that for $\delta_{\mu} \ll1$, i.e. for an RG scale $\mu$ far below $\Lambda_{\text{UV}}$, we can expand as \eqref{Eq:RG1Gamma}
\begin{align}\label{Eq:ApproxFixPoint}
    \lambda_Q^{1i}(\mu) \simeq \fp{\lambda_Q^{1i}}\left(1+O\left(\delta_{\mu}^{-2\gamma_Q^1}\right)\right) \,,
\end{align}
i.e. $\delta_{\mu}^{-2 \gamma_Q^1}$ controls corrections from the fixed point value in the IR. 

Assuming $\lambda_{Q}^{1i}$ reaches the fixed point soon after $\Lambda_{\text{UV}}$, we can replace \eqref{Eq:FP1} into \eqref{Eq:GenRGtop} and iterate for $\lambda_Q^{2i}$, once again neglecting $\lambda_{U_L,D_L}^{ai}$ entering the RG. We get
\begin{align}
\begin{cases}
    \frac{d \lambda_{Q}^{2 1}}{ d \log \mu} &= (\gamma^2_Q-\gamma^1_Q) \lambda_Q^{21}\\
    \frac{d \lambda_{Q}^{2 i}}{ d \log \mu} &= \gamma_Q^2 \lambda_Q^{2i}+\frac{ \lambda_Q^{2i}||\lambda_Q^2||^2}{\gst^2} \quad\quad i=2,3
\end{cases}
\end{align}
The off-diagonal mixing $\lambda_{Q}^{21}$ is exponentially suppressed, since $\gamma_Q^2-\gamma_Q^1>0$, while $||\lambda_Q^{2\,i=2,3}||$ approaches itself the fixed point (we assume this also happens close to $\Lambda_{\text{UV}}$ itself)
\begin{align}\label{Eq:FPQ2}
   \fp{\lambda_Q^{22}} = \gst \sqrt{-\gamma_Q^2}\,, && \fp{\lambda_Q^{23}}=0\,,
\end{align}
where we used the residual symmetries of the elementary quarks to set $\fp{\lambda_Q^{23}}$ to zero. 

Moving on to $\lambda_Q^{3i}$, it is straightforward to see from the associated RGs that $\lambda_Q^{3i=1,2} \rightarrow 0$ in the IR, since $\gamma_Q^3 - \gamma_Q^{1,2} > 0$, similarly to the fate of $\lambda_Q^{21}$ discussed above. On the other hand, the RG for $\lambda_Q^{33}$ is more involved, as it requires including the third most relevant interaction, corresponding to ${\cal O}_{U_L}^3$. We can verify that $\lambda_{U_L}^{3i=1,2} \rightarrow 0$ under RG flow, due to $\gamma_{U_L}^3 - \gamma_Q^{1,2} > 0$, thereby justifying our earlier neglect of these terms in the RGs for $\lambda_Q^{3i=1,2}$. As for $\lambda_{U_L}^{33}$, it reaches the fixed point
\begin{align}\label{Eq:FPU3}
    \fp{\lambda_{U_L}^{33}} = \gst \sqrt{-\gamma_{U_L}^3}\,,
\end{align}
which in turn drives $\lambda_Q^{33} \rightarrow 0$.

All the remaining entries of $\boldsymbol{\lambda}_{U_L,D_L}$, aside from $\lambda_{U_L}^{33}$, are suppressed by RG evolution. Approximating $\lambda_Q^{1\,i=1,2,3}$, $\lambda_Q^{2\,i=2,3}$, and $\lambda_{U_L}^{33}$ to be already close to their fixed points given in (\ref{Eq:FP1},~\ref{Eq:FPQ2},~\ref{Eq:FPU3}) near $\Lambda_{\text{UV}}$, by a straightforward but tedious evaluation of the RG, we find 
\begin{align}
\begin{aligned}\label{Eq:LHEndStageI}
    \boldsymbol{\lambda}_{U_L}(\mu)\simeq \begin{bmatrix}
        \delta_{\mu}^{\gamma_{U_L}^1-\gamma_Q^1}&&\delta_{\mu}^{\gamma_{U_L}^1-\gamma_Q^2} && \delta_{\mu}^{\gamma_{U_L}^1-\gamma_{U_L}^3}\\
        \delta_{\mu}^{\gamma_{U_L}^2-\gamma_Q^1}&&\delta_{\mu}^{\gamma_{U_L}^2-\gamma_Q^2} && \delta_{\mu}^{\gamma_{U_L}^2-\gamma_{U_L}^3}\\
        \delta_{\mu}^{-\gamma_Q^1}&&\delta_{\mu}^{-\gamma_Q^2} && \gst \sqrt{-\gamma_{U_L}^3}
    \end{bmatrix}\,,\\
    \boldsymbol{\lambda}_{D_L}(\mu)\simeq \begin{bmatrix}
        \delta_{\mu}^{\gamma_{D_L}^1-\gamma_Q^1}&&\delta_{\mu}^{\gamma_{D_L}^1-\gamma_Q^2} && \delta_{\mu}^{\gamma_{D_L}^1-\gamma_{U_L}^3}\\
        \delta_{\mu}^{\gamma_{D_L}^2-\gamma_Q^1}&&\delta_{\mu}^{\gamma_{D_L}^2-\gamma_Q^2} && \delta_{\mu}^{\gamma_{D_L}^2-\gamma_{U_L}^3}\\
        \delta_{\mu}^{\gamma_{D_L}^3-\gamma_Q^1}&&\delta_{\mu}^{\gamma_{D_L}^3-\gamma_Q^2} && \delta_{\mu}^{\gamma_{D_L}^3-\gamma_{U_L}^3}
    \end{bmatrix}\,,
    \end{aligned}
\end{align}
where we drop $O(1)$ factors deriving from the specific $\boldsymbol{\lambda}_{Q,U_L,D_L}(\Lambda_{\text{UV}})\sim O(1)$. 
We emphasize, again, that ultimately only the entries $\lambda_Q^{11},\lambda_Q^{22},\lambda_{U_L}^{33}$ are sizable in the IR while the others flow to zero, suppressed by the RG. 
This feature indeed justifies our initial (approximate) approach of solving independently for $\lambda_{Q}^{11}$, $\lambda_{Q}^{22}$ and $\lambda_{U_L}^{33}$ while basically neglecting other terms.

As explained in the main text the RG evolution of the RH mixings is not significant in Stage-I since we always have $1\lesssim\boldsymbol{\lambda}_{U_R,D_R}\lesssim\gst$.

We now turn to Stage-II. Flavor symmetries constrain the scaling dimensions the various operators:
\begin{align}\label{Eq:StageIIApp}
    \text{Stage-II}:&&\begin{cases}
        \gamma_{U_L}^{a=1,2,3}\equiv \gamma_{U_L}<0\quad
        \gamma_{D_L}^{a=1,2,3}\equiv \gamma_{D_L}>0\\
        \gamma_{U_R}^{a=1,2,3}\equiv \gamma_{U_R}<0\quad
        \gamma_{D_R}^{a=1,2,3}\equiv \gamma_{D_R}<0
    \end{cases}
\end{align}

Let us start with the RH mixings. By a rotation of the elementary $u^i,d^i$ and composite $\op_{U_R,D_R}^{a}$ we can bring $\boldsymbol{\lambda}_{U_R,D_R}(\Lfl)$ into diagonal form $\lambda_{U_R,D_R}^{a\neq i}(\Lfl)\equiv 0$. The RG evolution is than simply
\begin{align}
    \dfrac{d\lambda_{U_R}^{ii} }{d\log \mu} = \gamma_{U_R}\lambda_{U_R}^{ii} +\frac{(\lambda_{U_R}^{ii})^3}{\gst^2}\,,
\end{align}
for $i=1,2,3$. Again, $\boldsymbol{\lambda}_{U_R}$ reach the universal fixed point 
\begin{align}\label{Eq:IdentityUPMFV}
\fp{\boldsymbol{\lambda}_{U_R}}=\gst\sqrt{-\gamma_{U_R}}\,\mathbf{I}_3 \equiv \rho_{U_R} \mathbf{I}_3\,,
\end{align}
and analogously for the down-sector. 

The starting point of the RG for $\boldsymbol{\lambda}_{U_L,D_L}$ in Stage-II is
\begin{align}\label{Eq:LULFL}
    \lambda_{U_L}^{ai} (\Lfl) = c^{ab}_u \lambda_{U_L}^{bi}(\Lfl^+)\,,
\end{align}
for the up-sector, where $\lambda_{U_L}^{bi}(\Lfl^+)$ is \eqref{Eq:LHEndStageI} evaluated at $\mu=\Lfl$ and $c^{ab}_{u}$ accounts for the rearrangement of the degrees of freedom at the transition at $\Lfl$ (see main text). It is convenient to perform a rotation of the composite operators $\op_{U_L}^a$ in order to put $c_{u}^{ab}$ in lower triangular form. Then, given the hierarchical structure of $\boldsymbol{\lambda}_{U_L}(\Lfl^+)$, $\boldsymbol{\lambda}_{U_L}(\Lfl)$ in \eqref{Eq:LULFL} takes the same form of \eqref{Eq:LHEndStageI} up to irrelevant $O(1)$ factors. Note that the same rotation can be performed on $\op_{U_R}^a$, following the $U(3)_U$ symmetry, and on $u^i$, in order to maintain \eqref{Eq:IdentityUPMFV}. A similar consideration applies to the down-sector. Furthermore, all couplings in \eqref{Eq:LHEndStageI} are assumed to be rather small, with the only exception given by $\lambda_{U_L}^{33}$. We thus approximate the RG in Stage-II as 
\begin{align}
\begin{cases}
    \dfrac{d \lambda_{U_L}^{ai}}{ d \log \mu} = \, \gamma_{U_L}\lambda_{U_L}^{ai} + \lambda_{U_L}^{ai}\frac{(\lambda_{U_L}^{33})^2}{\gst^2} \delta^{i3}\quad a=1,2\\
    \dfrac{d \lambda_{U_L}^{ai}}{ d \log \mu} = \, \gamma_{U_L}\lambda_{U_L}^{ai} + \lambda_{U_L}^{ai}\frac{(\lambda_{U_L}^{33})^2}{\gst^2}\quad\quad a=3\\
    \dfrac{d \lambda_{D_L}^{ai}}{ d \log \mu} =  \gamma_{D_L}\lambda_{D_L}^{ai} + \lambda_{D_L}^{ai}\frac{(\lambda_{U_L}^{33})^2}{\gst^2} \delta^{i3}
\end{cases}
\end{align}
As usual, $\lambda_{U_L}^{33}$ is assumed to approach its fixed point $\lambda_{U_L}^{33}\sim \gst\sqrt{-\gamma_{U_L}}$ rapidly, while the remaining couplings are either rescaled by (non‐integer) powers of $\delta'_{\mu}=\mu/\Lambda_{\rm fl}$ or stay approximately constant. At $\mst$ we estimate
\begin{align}\label{Eq:LstageIIUp}
\boldsymbol{\lambda}_{U_L}(\mst) &\sim (\delta'_{\mst})^{\gamma_{U_L}}
\scriptsize
\begin{bmatrix}
\delta_{\Lfl}^{\gamma_{U_L}^1-\gamma_Q^1} & \delta_{\Lfl}^{\gamma_{U_L}^1-\gamma_Q^2} & \dfrac{\delta_{\Lfl}^{\gamma_{U_L}^1-\gamma_{U_L}^3}}{(\delta'_{\mst})^{\gamma_{U_L}}} \\
\delta_{\Lfl}^{\gamma_{U_L}^2-\gamma_Q^1} & \delta_{\Lfl}^{\gamma_{U_L}^2-\gamma_Q^2} & \dfrac{\delta_{\Lfl}^{\gamma_{U_L}^2-\gamma_{U_L}^3}}{(\delta'_{\mst})^{\gamma_{U_L}}} \\
\frac{\delta_{\Lfl}^{-\gamma_Q^1}}{(\delta'_{\mst})^{\gamma_{U_L}}} & \frac{\delta_{\Lfl}^{-\gamma_Q^2}}{(\delta'_{\mst})^{\gamma_{U_L}}} & \dfrac{\gst\,\sqrt{-\gamma_{U_L}}}{(\delta'_{\mst})^{\gamma_{U_L}}}
\end{bmatrix}
\normalsize
\end{align}
where we neglected $O(1)$ numbers in the matrix in brackets. A completely analogous expression holds for the down-sector, namely
\begin{align}\label{Eq:LstageIIDown}
\boldsymbol{\lambda}_{D_L}(\mst) &\sim (\delta'_{\mst})^{\gamma_{D_L}}
\scriptsize
\begin{bmatrix}
\delta_{\Lfl}^{\gamma_{D_L}^1 - \gamma_Q^1} & \delta_{\Lfl}^{\gamma_{D_L}^1 - \gamma_Q^2} & \dfrac{\delta_{\Lfl}^{\gamma_{D_L}^1 - \gamma_{U_L}^3}}{(\delta'_{\mst})^{\gamma_{U_L}}} \\
\delta_{\Lfl}^{\gamma_{D_L}^2 - \gamma_Q^1} & \delta_{\Lfl}^{\gamma_{D_L}^2 - \gamma_Q^2} & \dfrac{\delta_{\Lfl}^{\gamma_{D_L}^2 - \gamma_{U_L}^3}}{(\delta'_{\mst})^{\gamma_{U_L}}} \\
\delta_{\Lfl}^{\gamma_{D_L}^3 - \gamma_Q^1} & \delta_{\Lfl}^{\gamma_{D_L}^3 - \gamma_Q^2} & \dfrac{\delta_{\Lfl}^{\gamma_{D_L}^3 - \gamma_{U_L}^3}}{(\delta'_{\mst})^{\gamma_{U_L}}}
\end{bmatrix}
\normalsize
\,.
\end{align} 
Finally, as dictated by \eqref{Eq:MFVAns2}, (\ref{Eq:LstageIIUp},~\ref{Eq:LstageIIDown}) exhibit a hierarchical structure, similar to that of the Yukawa couplings in APC. However, in the up sector, both the left- and right-handed mixing angles are suppressed. In contrast, the left-handed mixing angles in the down sector are analogous to those in APC. Therefore, the $V_{\text{CKM}}$ mixing angles are dominated by the down-sector contribution in \eqref{Eq:LstageIIDown}, and are estimated to be of the order:
\begin{align}
    \theta_{12} &\sim \delta_{\Lfl}^{-\gamma_Q^1+\gamma_Q^2}\,, 
    & \theta_{23} &\sim \frac{\delta_{\Lfl}^{-\gamma_Q^2+\gamma_{U_L}^3}}{(\delta'_{\mst})^{-\gamma_{U_L}}}\,, 
    & \theta_{13} &\sim \frac{\delta_{\Lfl}^{-\gamma_Q^1+\gamma_{U_L}^3}}{(\delta'_{\mst})^{-\gamma_{U_L}}}\,.
\end{align}

We have checked that the above approximate analytic solution of the RG's for the various $\lambda$'s are in full agreement with a numerical evaluation of the full RG at $O(\lambda^3)$. Indeed, our benchmark points presented in \eqref{Eq:Benchmark} are based on such a complete analysis.

\section{A less symmetric model}\label{App:LessSModel}
In the main text we discussed in detail an UV realization of a maximally symmetric scenario (MFV). Less symmetric scenarios, yet not completely anarchic, are compelling options still compatible with a compositeness scale $\mst$ at the boundary of the (HL-)LHC reach $\sim O(5\,\TeV)$ \cite{Glioti:2024hye}(see Fig.~\ref{Fig:SummBounds}). 

We here briefly discuss how to generalize our construction to this class of models. For concreteness, we focus on the quark sector of the model with partial up-type universality, which constitutes a minimal departure from \eqref{Eq:MFVAns}. Namely, from the bottom up we consider the following EFT
\begin{align}\label{Eq:NMFVAns}
\begin{aligned}
    {\cal L}_{\rm EFT}=\frac{m_*^4}{g_*^2}\widehat{\cal L}_{\rm EFT}\Big(&\frac{\lambda^{Ai}_{U_L} q^i}{m_*^{3/2}},\frac{\lambda^{i}_{T_L} q^i}{m_*^{3/2}},\frac{\lambda^{ai}_{D_L} q^i}{m_*^{3/2}},\\
    &\frac{\rho_{U_R} u^A}{m_*^{3/2}},\frac{\rho_{T_R} t}{m_*^{3/2}},\frac{\rho_{D_R} d^a}{m_*^{3/2}},\ldots \Big)\,,
\end{aligned}
\end{align}
where $A=1,2$ and $a=1,2,3$. In the notation of \eqref{Eq:GenEFT}, the previous mixings are
\begin{align}\label{Eq:NMFVAnsStruct}
\begin{aligned}
\boldsymbol{\lambda}_{q} = 
\begin{bmatrix}
    \hat{\lambda}_{U_L}\\
    (\vec{\lambda}_{T_L})^T\\
    \boldsymbol\lambda_{D_L}
\end{bmatrix}
\, , \quad
\boldsymbol\lambda_{u} =
\left[
 \begin{array}{cc}
    \rho_{U_R} \, \mathbf{I}_2 & \mathbf{0}_{2 \times 1} \\ 
    \mathbf{0}_{1 \times 2} & \rho_{T_R} \\ 
    \multicolumn{2}{c}{\mathbf{I}_3}
  \end{array}
\right] ,\\
\quad 
\boldsymbol\lambda_{d} = 
\rho_{D_R}\,
\begin{bmatrix}
    \mathbf{0}_3 \\[6pt]
    \mathbf{I}_3
\end{bmatrix}\,, \quad \quad\quad \quad
\end{aligned}
\end{align}
with $\boldsymbol{\lambda}_{D_L}$ a $3 \times 3$ matrix, $\hat{\lambda}_{U_L}$ a $2 \times 3$ matrix, and $\vec{\lambda}_{T_L}$ a 3-vector. The global symmetry group is now $U(2)_U\times U(1)_T\times U(3)_D$, with $\hat{\lambda}_{U_L}$, $\vec{\lambda}_{T_L}$, and $\boldsymbol{\lambda}_{D_L}$ such that
\begin{align}
\begin{aligned}
    \lambda_{U_L}^{Ai}q^{i} &\in \bar{\mathbf{2}}_{U(2)_U}\,, &\quad u^A &\in \mathbf{2}_{U(2)_U}\,,\\ 
    \lambda_{D_L}^{ai}q^{i} &\in \bar{\mathbf{3}}_{U(3)_D}\,, &\quad d^a &\in \mathbf{3}_{U(3)_D}\,,
\end{aligned}
\end{align}
and $t$ and $\vec{\lambda}_{T_L} \cdot q$ transforming under $U(1)_T$ with opposite charges.

The down-sector reproduces the Yukawa as in \eqref{Eq:MFVAns2}, while the up-sector has
\begin{align}
   Y_u^{ia} \simeq \left[ \begin{array}{cc}
   c_u\dfrac{(\lambda_{U_L}^{Ai})^{\dagger}\rho_{U_R}}{\gst}&c_t\dfrac{\lambda_{T_L}^{i*}\rho_{\rho_R}}{\gst}\end{array} \right]\,,
\end{align}
with $c_{u,t}$ $O(1)$ Wilson coefficients.

We want to generate the previous ansatz dynamically as a result of RG evolution. For that it is straightforward to generalize the two-stages RG presented in the main text. Indeed, the starting point is exactly \eqref{Eq:MixPhaseI} with the additional constraint that  Stage-I is not completely anarchic but possesses a $U(1)_T$ symmetry. Under that the composites operators $\op_{T_L}\equiv \op_{U_L}^{a=3}$ and $\op_{T_R}\equiv \op_{U_R}^{a=3}$ have the same charge.  Aside from this, the RG evolution in Stage-I is completely analogous to what described in the main text and detailed in App.~\ref{App:ExtraDetails}. Indeed, the additional $U(1)_T$ symmetry is completely negligible in this first stage. The LH mixings evolve as (\ref{Eq:LstageIIUp},~\ref{Eq:LstageIIDown}), while the RH mixings are always $\lesssim \gst$.

In Stage-II, the flavor symmetries $U(2)_U \times U(3)_D$ are emergent. The $U(1)_T$ symmetry is assumed to be the same in both stages. The RG flow closely resembles that of MFV, with the key difference
\begin{align}
    \text{Stage-II:}&&\begin{cases}
        \gamma_{U_L}^{1,2}\equiv \gamma_{U_L}\quad \quad \gamma_{U_L}^3 \equiv \gamma_{T_L}\\
        \gamma_{U_R}^{1,2}\equiv \gamma_{U_R} \quad \quad \gamma_{U_R}^3 \equiv \gamma_{T_R}
    \end{cases}
\end{align}
as mandated by the $U(2)_U\times U(1)_T$ symmetry. We assume $\gamma_{U_R},\gamma_{T_R},\gamma_{T_L}\lesssim0$ while $\gamma_{U_L}\gtrsim0$, differently from the fully symmetric model. The $\lambda_{U_R}^{ai}$ approaches a non-universal fixed point
\begin{align}\label{Eq:NMFVRH}
    \fp{\lambda_{U_R}^{ai}} = \begin{pmatrix}
        \rho_U& 0 & 0 \\
        0 & \rho_U & 0 \\
        0 & 0 & \rho_T
    \end{pmatrix}=\gst\begin{pmatrix}
        \sqrt{-\gamma_U} & 0 & 0 \\
        0 & \sqrt{-\gamma_U} & 0 \\
        0 & 0 & \sqrt{-\gamma_T}
    \end{pmatrix},
\end{align}
matching \eqref{Eq:NMFVAnsStruct}. 

Regarding the LH mixings, the starting point of Stage-II is \eqref{Eq:LULFL}. The $U(1)_T$ symmetry constrains the structure of the anarchic matrix $c_u^{ab}$ entering in \eqref{Eq:LULFL}, enforcing the entries $c_u^{13} = c_u^{23} = c_u^{31} = c_u^{32} = 0$. Similarly to the MFV case, we can put the 2x2 block of $c_u^{ab}$ ($a,b =1,2$) in upper triangular form performing and $U(2)_U$ rotation on $\op_{U_L}^a$. This can be reabsorbed in the $O(1)$ factor of the hierarchical matrix in \eqref{Eq:LstageIIUp}. Note that the same rotation must be applied to $\op_{U_R}^a$ following the $U(2)_U$ symmetry and to $u^a$ to maintain the form of \eqref{Eq:NMFVRH}. The resulting RG flow is very similar to \eqref{Eq:LstageIIUp}. In particular we get
\begin{align}\label{Eq:LstageIIUpNMFV}
\boldsymbol{\lambda}_{U_L}^{ai}(\mst) &\sim (\delta'_{\mst})^{\gamma_{U_L}}
\scriptsize
\begin{bmatrix}
\delta_{\Lfl}^{\gamma_{U_L}^1-\gamma_Q^1} & \delta_{\Lfl}^{\gamma_{U_L}^1-\gamma_Q^2} & \dfrac{\delta_{\Lfl}^{\gamma_{U_L}^1-\gamma_{U_L}^3}}{(\delta'_{\mst})^{\gamma_{T_L}}} \\
\delta_{\Lfl}^{\gamma_{U_L}^2-\gamma_Q^1} & \delta_{\Lfl}^{\gamma_{U_L}^2-\gamma_Q^2} & \dfrac{\delta_{\Lfl}^{\gamma_{U_L}^2-\gamma_{U_L}^3}}{(\delta'_{\mst})^{\gamma_{T_L}}} \\
\frac{\delta_{\Lfl}^{-\gamma_Q^1}}{(\delta'_{\mst})^{\gamma_{U_L}}} & \frac{\delta_{\Lfl}^{-\gamma_Q^2}}{(\delta'_{\mst})^{\gamma_{U_L}}} & \dfrac{\gst\,\sqrt{-\gamma_{T_L}}}{(\delta'_{\mst})^{\gamma_{U_L}}}
\end{bmatrix}
\normalsize
\end{align}
and we can identify $\boldsymbol{\lambda}_{U_L}(\mst)= \left[ \hat{\lambda}_{U_L}^T(\mst),\vec{\lambda}_{T_L}^T (\mst)\right]^T$. The down-sector is completely equivalent to MFV and we get \eqref{Eq:LstageIIDown} with the replacement $\gamma_{U_L}\rightarrow \gamma_{T_L}$.

The benchmark values $\mst= 5 \TeV$, $\gst=3$, not excluded by present constraints, with $\Lfl= 10^6 \TeV$, $\Lambda_{\text{UV}}=m_{\text{Planck}}$ and 
\begin{align}\label{Eq:BenchmarkNMFV}
&\begin{aligned}
\text{Stage-I: }\quad\quad
\gamma_{U_L}^a &\sim (0.1,\, 0.05,-0.05),\\ 
-\gamma_Q^a &\sim (0.35,\,0.25,\,0.05),\,\\
\gamma_{D_L}^a &\sim (0.1,\, 0.05,\, 0.05),\\
-\gamma_{U_R,D_R}^a &\sim O(0.1)
\end{aligned}\notag\\
&\begin{aligned}
\text{Stage-II: }\quad
&\gamma_{U_L} \sim 0.05,\,
\gamma_{U_R} \sim -0.15\\
&\gamma_{D_L} \sim 0.05,\, 
\gamma_{D_R} \sim -0.5\\
&\gamma_{T_L}\sim-0.07,\, \gamma_{T_R}\sim-1
\end{aligned}
\end{align}
give quark masses and mixings in the ballpark of the SM values. Furthermore, notice that $\gamma_{T_R} = -1$ corresponds to the limit where $\op_{T_R}$ becomes free. It can be shown \cite{Contino:2003ve} that this is dual to having the top $t_R$ fully composite.

\section{Experimental Constraints}\label{App:ExpCon}

Here we discuss the main low-energy experimental constraints on the models presented in the main text and App~\ref{App:LessSModel}. A comprehensive analysis is provided in \cite{Glioti:2024hye}, from which all the results quoted below are taken. We organize the discussion into two parts: quark-sector and lepton-sector constraints. All the quoted bound refer to exclusion of new physics at 95\% CL. A summary of the resulting constraints on the $(\mst,\,\gst)$-plane are shown in Fig.~\ref{Fig:SummBounds}.

\textit{\textbf{Quarks--}} For convenience, we present the relevant constraints on the Lagrangian in \eqref{Eq:NMFVAns}. These can be readily translated to \eqref{Eq:MFVAns} by the substitution $\rho_T \rightarrow \rho_U$. It is further convenient to normalize $\rho_{U,T,D}$ in units of $\gst$:
\begin{align}
    \rho_U &\equiv \eu\, \gst\,, \qquad
    \rho_T \equiv \et\, \gst\,, \qquad
    \rho_D \equiv \ed\, \gst\,,
\end{align}
with $\eu,\et \leq 1$.

The models feature sizably composite light right-handed quarks, controlled by $\eu$ and $\ed$ ($\rho_{U}$ and $\rho_D$). These are tested by measurements of four-fermion interactions in dijet events at the LHC, which require (at 95\% CL)
\begin{align}\label{Eq:dijet}
    \frac{m_*}{\gst\, \eu^2} &\gtrsim 4.8\text{--}7.8\ (8.2\text{--}13)\,\TeV\,, \\
    \frac{m_*}{\gst\, \ed^2} &\gtrsim 3.0\text{--}3.6\ (5.1\text{--}5.8)\,\TeV\,,
\end{align}
where the numbers in parentheses refer to the projected sensitivity at the end of Run~3 ($300\,\text{fb}^{-1}$). In each case, the two values correspond to destructive and constructive interference with the Standard Model.\footnote{Note, however, that integrating out an $s$-channel heavy spin-1 resonance results only in destructive interference.}

Among flavor transitions, the most significant effects are $\Delta F = 1$ in $b \rightarrow s$ transitions mediated by a $Z$ exchange. These are tightly constrained by the decay $B_s \rightarrow \mu^+ \mu^-$, which leads to the bound 
\begin{align}\label{Eq:DF1}
    m_* \gtrsim \frac{6.5\text{--}8.3}{\et}\,\TeV\,,
\end{align}
at 95\% CL.

In the model with partial up-type universality prefers $\et\sim 1$. In contrast, in the model with complete up-type universality (MFV) $\et = \eu$, and thus the lowest allowed $\mst$ for any $\gst$ is obtained combining \eqref{Eq:DF1} with \eqref{Eq:dijet} (see Fig.~\ref{Fig:SummBounds}).

The effect in \eqref{Eq:DF1} can be suppressed by imposing custodial protection of the $Z$ couplings, as proposed in \cite{Agashe:2006at}. In such scenarios, $\Delta F = 1$ effects become effectively negligible, and the dominant flavor transitions are associated with $B_d$ oscillations, leading to the bound
\begin{align}\label{Eq:DF2}
    m_* \gtrsim \frac{6.6}{\gst\, \varepsilon_{t}^2}\,\TeV\,,
\end{align}
at 95\%CL.
Again, for models with full RH universality $\et=\eu$ so that \eqref{Eq:DF2} must be combined with \eqref{Eq:dijet}. On the contrary the model of \eqref{Eq:NMFVAns} prefers $\et\sim1$.

In Fig.~\ref{Fig:SummBounds} we show the sensitivity reaches for models with full and partial universality, with and without custodial protection of the $Z$ couplings, optimized on $\varepsilon_{u,t}$ to allow the lowest $\mst$. For any bound quoted above it is conservatively considered the stronger side of the interval. The figure also show universal constraint (gray region), arising from the measurement of bosonic operators and assuming the strong sector is protected by large effects in $\hat{T}$ (see \cite{Agashe:2003zs}). The black, dashed line shows the direct LHC reach on spin-1 resonances kinetically mixed with the SM gauge bosons.

In the main text, in \eqref{Eq:Benchmark}, we consider the following benchmark point:
\begin{align}\label{Eq:BenchmarkApp}
    m_* &= 10\,\TeV\,, \,\,\,\gst = 3\,, \,\,\, \eu= 0.7,\,\,\,\ed=0.8\,,
\end{align}
which is at the edge of present constraints on MFV without custodial protection and well within that in the opposite case.

As a more optimistic option, in \eqref{Eq:BenchmarkNMFV} we also consider the benchmark point
\begin{align}\label{Eq:BenchmarkNMFVApp}
\begin{aligned}
    m_* &\sim 5\,\TeV\,, \qquad \gst \sim 3\,, \qquad \et \sim 1\,,  \\ &\qquad\eu\sim 0.4\,, \qquad \ed\sim 0.7\,,
\end{aligned}
\end{align}
which is at the boundary of the current LHC direct reach yet well within the indirect flavor bounds in partially symmetric models with custodial protection.

Because the RH flavor universality in \eqref{Eq:MFVAnsStruct} arises dynamically from the RG, there will be deviations at $\mst$, which will results in flavor violation at $\mst$. Consider for instance the down-sector, where corrections to \eqref{Eq:IdentityUPMFV} at $\mst$ are at most of the order
\begin{align}
    \lambda_{D_R}^{ai}(\mst) = \rho_{D_R} \mathbf{I}_3 \left(1+O\left(\frac{\mst}{\Lfl}\right)^{-2 \gamma_{D_R}}\right)\,,
\end{align}
as can be seen from \eqref{Eq:ApproxFixPoint}. Bounds on $C_1$ for $\Delta F=2$ transitions in the $sd$-system require \cite{Bona:2024bue}
\begin{align}
    \frac{\gst^2\ed^4}{\mst^2 }\left( \frac{\mst}{\Lfl}\right)^{-4 \ed^2}\lesssim (10^{4}-10^{5} \TeV)^{-2}\,,
\end{align}
where, recall $\ed^2 = - \gamma_{D_R}$. Assuming $\Lfl = 10^6$, the previous constraint is compatible with both (\ref{Eq:BenchmarkApp},~\ref{Eq:BenchmarkNMFVApp}). A similar expression holds for the up-sector current bounds from the $uc$-system $\Delta F=2$ \cite{Bona:2024bue} are well compatible with \eqref{Eq:BenchmarkApp} and marginally compatible with \eqref{Eq:BenchmarkNMFVApp}. Furthermore, notice that these effects can be simply suppressed by pushing $\Lfl$ further into the UV.

\textit{\textbf{Leptons--}}In the lepton sector, \eqref{Eq:MFVLeptEFT}, via \eqref{Eq:LeptYuk}, leads to no dangerous flavor transition in the approximation of neglecting the very small $\tilde{\mathcal{Y}}_{\nu}$. Yet, the compositeness of right-handed leptons, i.e. the parameter $\varepsilon_d \equiv \tilde{\rho}_D/\gst$ in \eqref{Eq:MFVLeptEFT}, is bounded by LEP data. Assuming a rough $10^{-3}$ precision on $Z$ couplings, we can estimate 
\begin{align}\label{Eq:BoundLEP}
\frac{\gst^2}{\mst^2}v^2{\varepsilon}_{e}^2 \lesssim \,10^{-3} &&\Longrightarrow &&\mst\gtrsim 8 \gst \varepsilon_{e}\TeV\,. 
\end{align}

Simultaneous quark and lepton compositeness can also lead to effects in high-energy di-lepton Drell–Yan measurements at the LHC. In particular, the projected sensitivity of the LHC with 300~$\text{fb}^{-1}$ for the dimension-6 operator $(\bar{u}\gamma^{\mu}u)(\bar{e}\gamma_{\mu}e)$~\cite{Panico:2021vav}, at 95\% CL implies
\begin{align}\label{Eq:BounDY}
    \frac{\gst^2 \eu^2 \varepsilon_e^2}{\mst^2} \lesssim 4 \times10^{-3}~\TeV^{-2}\,.
\end{align}

The benchmark point in~\eqref{Eq:BenchmarkApp} is compatible with both~\eqref{Eq:BoundLEP} and~\eqref{Eq:BounDY}, provided that $\varepsilon_e \lesssim 0.4$.

\begin{figure}[t!]
    \centering
    \includegraphics[width=1\linewidth]{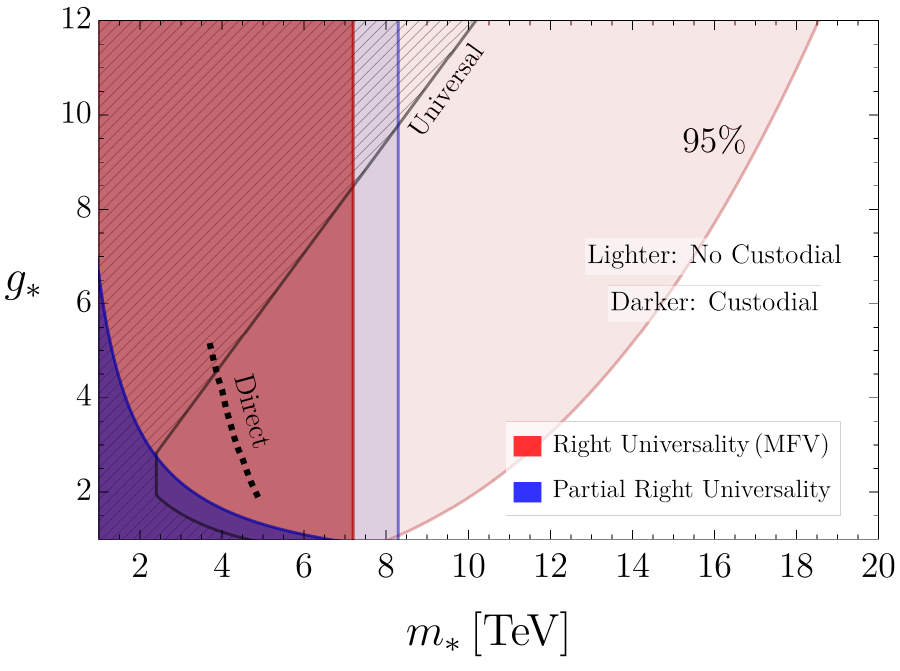}
    \caption{Main sensitivity bounds for the flavor symmetric models presented in the text.}
    \label{Fig:SummBounds}
\end{figure}

\bibliography{bibliography} 

\end{document}